%% file: main.tex
\documentclass[11pt,a4paper]{article}
\usepackage[T1]{fontenc}
\usepackage{lmodern}
\usepackage{amsmath,amssymb,amsthm,mathtools}
\usepackage[a4paper,margin=1.05in,includefoot]{geometry}
\usepackage{enumitem,booktabs}
\usepackage[ruled,vlined,linesnumbered]{algorithm2e}
\usepackage[protrusion=false]{microtype}
\usepackage[colorlinks=true,allcolors=blue]{hyperref}
\newtheorem{theorem}{Theorem}[section]
\newtheorem{lemma}[theorem]{Lemma}
\newtheorem{proposition}[theorem]{Proposition}
\newtheorem{corollary}[theorem]{Corollary}
\theoremstyle{definition}
\newtheorem{definition}[theorem]{Definition}
\newtheorem{example}[theorem]{Example}
\newcommand{\sig}{\sigma}

\newcommand{\calT}{\mathcal T}
\newcommand{\calD}{\mathcal D}
\newcommand{\calL}{\mathcal L}
\newcommand{\dhat}{\widehat d}
\newcommand{\dist}{\operatorname{dist}}
\newcommand{\J}{\operatorname{J}}
\newcommand{\up}{\operatorname{up}}
\newcommand{\bag}{\operatorname{bag}}
\newcommand{\owner}{\operatorname{owner}}
\newcommand{\disc}{\operatorname{disc}}
\newcommand{\size}[1]{\lvert #1\rvert}
\newcommand{\Admit}{\textnormal{\textsc{Admit}}}
\newcommand{\Search}{\textnormal{\textsc{Search}}}
\newcommand{\Prepare}{\textnormal{\textsc{Prepare}}}
\newcommand{\Learn}{\textnormal{\textsc{Learn}}}
\newcommand{\Add}{\textnormal{\textsc{Add}}}
\newcommand{\Select}{\textnormal{\textsc{Select}}}
\newcommand{\Solve}{\textnormal{\textsc{TerminalSolve}}}
\newcommand{\Interior}{\textnormal{\textsc{Interior}}}
\SetKwInput{KwIn}{Input}
\SetKwInput{KwOut}{Output}
\SetKwProg{Proc}{procedure}{}{}
\SetKwComment{Rem}{$\triangleright$\ }{}

\SetCommentSty{remsty}
\DontPrintSemicolon
\SetAlFnt{\small}
\newcommand{\DocumentTitle}{Faster Directed SSSP on Digraphs with Random Sparse Supports or Few DFS Exceptions}
\hypersetup{
  pdftitle={\DocumentTitle},
  pdfauthor={Valery (Binyamin) Grishin, Aryeh Lev Zabokritskiy (Yohananov)},
  pdflang={en}
}
\title{\DocumentTitle}
\author{
  \normalsize Valery (Binyamin) Grishin\\[0.5em]
  \small Department of Computer Science, Tel-Hai University of Kiryat Shmona\\
  \small and the Galilee, Kiryat Shmona, Israel\\[0.25em]
  \small\href{mailto:valerig.tech@gmail.com}{\texttt{valerig.tech@gmail.com}}\\[2em]
  \normalsize Aryeh Lev Zabokritskiy (Yohananov)\\[0.5em]
  \small Department of Computer Science, MIGAL--Galilee Research Institute,\\
  \small Kiryat Shmona, Israel, and Department of Computer Science, Tel-Hai\\
  \small University of Kiryat Shmona and the Galilee, Kiryat Shmona, Israel\\[0.25em]
  \small\href{mailto:yuhanalev@telhai.ac.il}{\texttt{yuhanalev@telhai.ac.il}}\\[0.5em]
  \small ORCID: \href{https://orcid.org/0000-0003-3151-6192}{0000-0003-3151-6192}
}
\date{}
\begin{document}
{\hypersetup{urlcolor=black}\maketitle}
\input{opening}
\input{model}
\input{sparse}
\input{intermediate}
\input{conclusion}
\appendix
\input{import}

\input{references}
\end{document}

%% file: opening.tex
\begin{abstract}
We study exact single-source shortest paths (SSSP) in directed graphs with $n$ vertices, $m$ arcs and nonnegative real weights, in the comparison-addition model. This paper continues our earlier work, which gave a deterministic algorithm that takes $O(n\sqrt{\log n}\,(\log\log n)^{1/4})$ time when $m=O(n)$ and $O(n\sqrt{\log n}\,(\log\log n)^{5/8})$ time when $m=\Theta(n\sqrt{\log\log n})$. Our two deterministic algorithms compute the distances on every input and take $O(n\sqrt{\log n})$ time on two families of inputs, one per density. The support of an input is its underlying simple undirected graph. When $m=O(n)$, the first algorithm takes $O(n\sqrt{\log n})$ time if no two simple cycles of the support with fewer than $P$ vertices share a vertex, where $P=\Theta(\sqrt{\log n})$ depends only on the support. For fixed $c>0$, a uniformly random graph with $n$ vertices and $\lfloor cn\rfloor$ edges has this property with probability at least $1-n^{-1+o(1)}$, and the bound then holds for every choice of one or both directions of each edge, all weights and every source. The second family is defined through the depth-first search tree of the support that starts at the source and visits neighbours in identifier order. For fixed $k\ge1$, call a set $Y$ of exceptional vertices $k$-feasible if, for each nonroot vertex, at most $k$ of its proper ancestors are joined to its subtree by arcs whose endpoints avoid $Y$ and whose underlying edges lie outside the DFS tree. Given $k$ but not such a set, the second algorithm takes, for every $k$-feasible set $Y$ of total out-degree $d^+(Y)$, time $O_k\bigl(n+m+(1+\size Y)\log n+\sqrt{(n+m)\,d^+(Y)\log n}\bigr)$. When $m=\Theta(n\sqrt{\log\log n})$, this is $O_k(n\sqrt{\log n})$ if some $k$-feasible set has $O(n/\sqrt{\log n})$ vertices and total out-degree $O(n/\sqrt{\log\log n})$.
\end{abstract}

\section{Introduction}

Consider a directed graph with $n$ vertices, $m$ arcs and nonnegative real arc weights. The single-source shortest path problem (SSSP) asks for the distances from a source $s$ to all vertices. We work in the comparison-addition model, as in~\cite{DMMSY2025,DMSY2026}: the weights are real numbers that are only added and compared.

Dijkstra's algorithm with Fibonacci heaps solves the problem in $O(m+n\log n)$ time~\mbox{\cite{Dijkstra1959,FredmanTarjan1987}}. Duan, Mao, Mao, Shu and Yin~\cite{DMMSY2025} gave a deterministic algorithm with running time $O(n+m\log^{2/3}n)$, including the cost of reporting distances to unreachable vertices. Duan, Mao, Shu and Yin~\cite{DMSY2026} obtained the deterministic bound $O(n+m\sqrt{\log n}+\sqrt{mn\log n\log\log n})$. In our previous paper~\cite{GZ}, we improved this bound for sparse inputs and for a range of intermediate densities. Our earlier algorithm takes
\begin{equation}\label{eq:gz}
\begin{aligned}
 &O\bigl(n\sqrt{\log n}\,(\log\log n)^{1/4}\bigr)&&\text{time for }m=O(n),\\
 &O\bigl(n\sqrt{\log n}\,(\log\log n)^{5/8}\bigr)&&\text{time for }m=\Theta\bigl(n\sqrt{\log\log n}\bigr).
\end{aligned}
\end{equation}
We ask for which inputs the factors $(\log\log n)^{1/4}$ and $(\log\log n)^{5/8}$ can be removed, and we give one main family of inputs and one algorithm for each of the two densities. Both algorithms are deterministic and compute the exact distances and a shortest-path tree on every input; the hypotheses below concern only their running time.

Our first guarantee is for random supports. The \emph{support} of a directed graph is the simple undirected graph in which two distinct vertices are adjacent when an arc joins them. An adversary who knows the support may choose the directions, the weights and the source.

\begin{theorem}\label{thm:typical-sparse}
There is a deterministic algorithm for exact SSSP with the following property. Fix $c>0$, and choose a simple graph uniformly at random among the graphs with vertex set $\{1,\dots,n\}$ and $\lfloor cn\rfloor$ edges. With probability at least $1-n^{-1+o(1)}$, the algorithm runs in $O_c(n\sqrt{\log n})$ time on every directed graph obtained by replacing each edge by one or both of its orientations, for all nonnegative real weights and every source. The expectation, over the random graph, of the maximum running time over these inputs is also $O_c(n\sqrt{\log n})$.
\end{theorem}

The theorem follows from a bound for individual inputs. When $m=O(n)$, the algorithm takes $O(n\sqrt{\log n})$ time on every input in which no two simple cycles of the support with fewer than $P$ vertices share a vertex, where $P=\Theta(\sqrt{\log n})$ is computed by the algorithm from the vertex degrees of the support and is at most $\sqrt{\log_2 n}$ for $n\ge2^{16}$. On every input with $m=O(n)$ it takes $O(n\log n)$ time.

Our second result holds at every density. Fix the depth-first search tree of the support that starts at $s$ and visits neighbours in increasing order of their identifiers. A \emph{chord} is an arc between two distinct vertices of the weak component of $s$ that are not joined by a tree edge; one of its endpoints, the \emph{upper} one, is a proper ancestor of the other, the \emph{lower} one. A chord \emph{crosses the cut above} a vertex $v\ne s$ if its lower endpoint is $v$ or a descendant of $v$ and its upper endpoint is a proper ancestor of $v$. For an integer $k\ge1$, a set $Y$ of vertices is \emph{$k$-feasible} if, for every $v\ne s$, the chords with no endpoint in $Y$ that cross the cut above $v$ have at most $k$ distinct upper endpoints. Let $d^+(Y)$ be the number of arcs whose tail is in $Y$, parallel arcs counted once.

\begin{theorem}\label{thm:dfs-exceptions}
For every fixed $k\ge1$ there is a deterministic algorithm that, given a directed graph with nonnegative real weights and a source $s$, computes the distances from $s$ and a shortest-path tree in $O_k(n+m)$ space and in time
\[
 O_k\Bigl(n+m+\min_{Y}\Bigl[(1+\size Y)\log n+\sqrt{(n+m)\,d^+(Y)\log n}\,\Bigr]\Bigr),
\]
where the minimum is over the $k$-feasible sets $Y$ of the weak component of $s$.
\end{theorem}

The algorithm receives only the graph, the source and $k$; it selects its own $k$-feasible set of exceptional vertices and pays for the selection within the stated bound. For $m=\Theta(n\sqrt{\log\log n})$ the bound is $O_k(n\sqrt{\log n})$ whenever some $k$-feasible set $Y$ has $\size Y=O(n/\sqrt{\log n})$ and $d^+(Y)=O(n/\sqrt{\log\log n})$. There are inputs of this density, with strongly connected graphs, on which the bound is $O_k(n+m)$ for every $k\ge2$, while no fixed $k$ makes the empty set $k$-feasible in all of them. For $m=O(n)$ the bound is $O_k(n\sqrt{\log n})$ whenever some $k$-feasible set $Y$ has $\size Y=O(n/\sqrt{\log n})$, since $d^+(Y)\le m$.

We do not improve the bounds \eqref{eq:gz} for general inputs. Either of our algorithms can be run in alternation with that of~\cite{GZ}, one step of each in turn, until one of them stops; this takes at most twice the smaller of the two running times, so the bounds \eqref{eq:gz} are kept on all inputs. The algorithm of Theorem~\ref{thm:typical-sparse} runs, only on very large inputs, the recursive search of~\cite{GZ} with one of its steps replaced, and Dijkstra's algorithm otherwise; its analysis uses results proved in~\cite{GZ}. Theorem~\ref{thm:dfs-exceptions} uses~\cite{GZ} only for the conventions of the model and for the standard construction of a shortest-path tree from the distances.

Haeupler, Hlad\'ik, Rozho\v{n}, Tarjan and T\v{e}tek~\cite{HHRTT2024} showed that Dijkstra's algorithm with a suitable heap is universally optimal for the problem of ordering the vertices by their distance from the source; the bounds above are for computing the distances. Average-case analyses of SSSP take the arc weights to be random, and linear expected time is known in that setting~\cite{Goldberg2008,Meyer2003}. In Theorem~\ref{thm:typical-sparse} only the support is random, and the weights are arbitrary. For structured inputs, a shortest-path tree of a directed graph of constant treewidth can be computed in linear time~\cite{ChaudhuriZaroliagis2000}, and for nearly acyclic directed graphs there are algorithms whose running time depends on a parameter that is small for such graphs~\cite{SaundersTakaoka2003,Takaoka1998}. The bound of Theorem~\ref{thm:dfs-exceptions} is of this kind, with the minimum over the $k$-feasible sets $Y$ as its parameter: if the empty set is $k$-feasible, the weak component of $s$ has treewidth at most $k+1$ and the bound is $O_k(n+m)$.

Section~\ref{sec:model} fixes the model. Section~\ref{sec:sparse} gives the sparse algorithm and proves Theorem~\ref{thm:typical-sparse} together with the bounds for individual inputs. Section~\ref{sec:intermediate} proves Theorem~\ref{thm:dfs-exceptions} and gives the strongly connected examples. Section~\ref{sec:conclusion} states what remains open.

%% file: model.tex
\section{Model and conventions}\label{sec:model}

We use the comparison-addition model with the conventions of~\cite[Section~3]{GZ}: arc weights and path lengths are real numbers that are only added and compared, in unit time, and vertex identifiers, counters and pointers are integers of $O(\log(n+m+2))$ bits with unit-cost comparisons, additions, subtractions, shifts and array accesses; other integer operations are computed from these and charged as in~\cite[Section~3 and Appendix~F]{GZ}. For a nonnegative integer $z$ we write $\J(z)\coloneqq\lceil\log_2(z+2)\rceil$, as in~\cite{GZ}. A subscript on $O$ lists the parameters on which the implied constant depends.

The \emph{support} of a directed graph is the simple undirected graph on its vertices in which two distinct vertices are adjacent when an arc joins them, in either direction. A graph is \emph{simple} if it has no loops and at most one arc for each ordered pair of vertices. Both algorithms make the input simple: they remove the loops and keep one arc of least weight for each ordered pair of vertices. Radix sorting the arcs by the identifiers of their endpoints does this in $O(n+m)$ time. Throughout, $n$ and $m$ count the vertices and arcs of the input, before this step.

Both algorithms compute the distances $\dist(v)$ from the source $s$ to all vertices and then a shortest-path tree. An arc $(u,v)$ of weight $w$ is \emph{tight} if $\dist(u)$ is finite and $\dist(u)+w=\dist(v)$. A traversal of the tight arcs from $s$ that records, for every vertex, the arc by which it is first reached yields a shortest-path tree in $O(n+m)$ time, also when the graph has cycles of weight zero~\cite[Section~5.1]{GZ}. Both algorithms end with this traversal.

%% file: sparse.tex
\section{Sparse supports}\label{sec:sparse}

This section proves Theorem~\ref{thm:typical-sparse}. The algorithm is Algorithm~\ref{alg:driver}. It replaces the input by a graph of bounded degree and runs the recursive search of~\cite{GZ} on it, with one step replaced: the step that groups the sources of a recursive call. In~\cite{GZ} this step runs bounded shortest-path searches from the sources. Here every source claims vertices by a depth-first search, at constant cost per claimed vertex, and the claimed vertices pay for the first worklist entry of each group. A small connected component of the claimed vertices cannot pay for its group in this way. The step then retires a source of the component, which the analysis of~\cite{GZ} pays for, or it finds a cycle with few vertices in the component and stores a certificate for this cycle at one of its arcs, unless that arc already holds a fixed number $b$ of certificates. A stored certificate keeps this cycle from being found again at this arc. Theorem~\ref{thm:short-cycles} bounds the running time on every input whose short cycles are disjoint, and Theorem~\ref{thm:typical-sparse} follows by counting the short cycles of a random support.

\subsection{The algorithm and the short-cycle bound}\label{subsec:driver}

The input is a directed graph $H$ with $n$ vertices, $m$ arcs, nonnegative real arc weights and a source $s$. The algorithm first makes $H$ simple (Section~\ref{sec:model}).

The recursive search needs bounded degrees. Replace each vertex $x$ that has $\delta(x)\ge2$ neighbours in the support by $\delta(x)$ new vertices, its \emph{ports}, joined into a path by arcs of weight zero in both directions, and assign one port to each neighbour of $x$; the order of the ports along the path is arbitrary. A vertex with at most one neighbour is its own port. For adjacent $x$ and $y$, every arc of $H$ between $x$ and $y$ becomes an arc between the port of $x$ assigned to $y$ and the port of $y$ assigned to $x$, with the same direction and weight. The result is the \emph{routing graph} $G$, and $s'$ denotes a port of $s$. It differs from the routing graph of~\cite[Section~5.1]{GZ}, which, for a degree bound $D\ge3$ chosen there, gives a vertex one port for each segment of at most $D-2$ incidences and joins its ports into a directed cycle. Here every neighbour has its own port and the ports form a path, so that the ports of one vertex create no cycle in the support of $G$ (Section~\ref{subsec:cycles}). The graph $G$ is simple, has $N\le n+2m$ vertices and $M\le3N$ arcs, and its incoming and outgoing degrees are at most $3$, because a port has at most two neighbours on its path and one outside it. Every walk of $H$ lifts to a walk of $G$ of the same length that moves along the paths of ports at no cost, and every walk of $G$ projects to a walk of $H$ of the same length. Hence the distance from $s'$ to each port of $x$ in $G$ is the distance from $s$ to $x$ in $H$.

The search is recursive, and every call has a level $j\ge0$. A call of level $j$ completes vertices in increasing order of their distance from $s'$ until it has completed at least $H_j$ of them or all that it was asked for, and for $j\ge1$ it does so through calls of level $j-1$. The capacity $H_j$ grows by the factor $2^T$ from one level to the next, where $T$ is the \emph{scale} of the search. A larger scale gives fewer levels but more expensive operations on the worklists, in which the calls keep their pending vertices. Let $L\coloneqq\J(N)$. The scale $T$, the number $P$ of vertices that one source may own in the preparation step of a call (Section~\ref{subsec:prepare}), and the top level $e$ are
\begin{equation}\label{eq:scales}
 T\coloneqq\max\{4096,\lceil\sqrt L\,\rceil\},\qquad P\coloneqq\lfloor T/1024\rfloor,\qquad e\coloneqq\lceil\log_2(N)/T\rceil .
\end{equation}
A call at level $j$ has at most $A_j$ sources, pulls at most $M_j$ entries from its worklist at a time, and stops once its output has at least $H_j$ vertices, where
\[
 A_j\coloneqq T^2 2^{jT},\qquad M_0\coloneqq 1,\qquad M_j\coloneqq 16T\,2^{(j-1)T}\quad(j\ge1),\qquad H_j\coloneqq T^3 2^{jT}
\]
are the capacities of~\cite[(11)]{GZ} at these scales. Since $H_e>N$, the capacity $H_e$ does not stop the call at the top level $e$. The search is run only if the following \emph{guards} of~\cite[Section~5.2]{GZ} hold, where $D$ is a bound on the incoming and outgoing degrees of $G$:
\begin{equation}\label{eq:guards}
\begin{gathered}
 4096\le T\le L,\qquad T^2\le N,\qquad 2\le P\le N,\qquad 16(3P+1)\le T,\\
 2M_j+A_j+2DH_j\le M_j2^{2T}\quad(j\ge1).
\end{gathered}
\end{equation}
For $D=3$ the guards hold if and only if $L\ge4096$; the proof of Theorem~\ref{thm:short-cycles} verifies this. The search therefore runs only when $N+2>2^{4095}$; on smaller inputs Algorithm~\ref{alg:driver} runs Dijkstra's algorithm.

Algorithm~\ref{alg:driver} runs the procedure $\Search$, which maintains a label $\dhat(v)$ for every vertex $v$ of $G$, a walk from $s'$ to $v$ or $\infty$. The integer $b$ bounds the number of certificates that the search stores at one arc (Section~\ref{subsec:prepare}). It is a constant, and $b=1$ suffices for Theorem~\ref{thm:typical-sparse}.

\begin{algorithm}[htbp]
\caption{Shortest paths on sparse supports}\label{alg:driver}
\KwIn{directed graph $H$ with nonnegative real arc weights; source $s$; an integer $b\ge1$}
\KwOut{the distances from $s$ in $H$ and a shortest-path tree}
remove the loops of $H$; for each ordered pair of vertices keep one arc of least weight\;
$(G,s')\gets$ the routing graph of $H$ and a port of $s$\;\label{line:routing}
$N\gets\size{V(G)}$;\quad $L\gets\J(N)$;\quad compute $T$, $P$ and $e$ by \eqref{eq:scales}\;\label{line:scales}
\If{a guard in \eqref{eq:guards} fails for $D=3$}{\label{line:guardtest}
  \Return the output of Dijkstra's algorithm with a Fibonacci heap~\cite{FredmanTarjan1987} on $H$\;\label{line:fallback}
}
$\dhat(s')\gets$ the walk without arcs;\quad $\dhat(v)\gets\infty$ for every other vertex $v$ of $G$\;
the certificate list of every arc of $G$ $\gets$ empty\Rem*{it will hold at most $b$ certificates}
$\Search(+\infty,\{s'\},e)$\Rem*{Algorithm~\ref{alg:search}}\label{line:topcall}
\lForEach{vertex $x$ of $H$}{$\dist(x)\gets$ the length of $\dhat(x')$ for a port $x'$ of $x$}
\Return $\dist$ and the tree of first discoveries along the tight arcs of $H$\Rem*{Section~\ref{sec:model}}
\end{algorithm}

\begin{theorem}\label{thm:short-cycles}
For every fixed $b$, Algorithm~\ref{alg:driver} returns the distances from $s$ in $H$ in $O_b(n+m+NL)$ time. If the simple cycles with fewer than $P$ vertices in the support of $H$ are pairwise vertex-disjoint, its running time is $O_b(n+m+N\sqrt L)$.
\end{theorem}

Sections~\ref{subsec:search}--\ref{subsec:cycles} prove Theorem~\ref{thm:short-cycles}. We first derive Theorem~\ref{thm:typical-sparse} from it.

\begin{proof}[Proof of Theorem~\ref{thm:typical-sparse}]
Use Algorithm~\ref{alg:driver} with $b=1$. Fix $c>0$ and write $a\coloneqq\lfloor cn\rfloor$. A directed graph obtained from a support with $a$ edges has $m\le2a$ arcs. Its routing graph has one vertex for every vertex of degree at most one and $\delta(x)$ vertices for every other vertex $x$, so $N\le n+2a\le(1+2c)n$. Then $L\le\log_2n+O_c(1)$, and $P\le T/1024$ gives $P\le p\coloneqq\lceil\sqrt{\log_2n}\,\rceil$ for all $n\ge n_0(c)$. Let $\mathcal E$ be the event that two distinct simple cycles with at most $p$ vertices in the random support share a vertex.

\emph{The probability of $\mathcal E$.} The union of two such cycles is a connected graph with $\ell\le2p$ vertices that contains two distinct cycles, so it has at least $\ell+1$ edges. The support is a uniformly random set of $a$ of the $\binom n2$ pairs of vertices, so any $i$ given pairs are all edges of the support with probability at most $\bigl(a/\binom n2\bigr)^i\le(2c/(n-1))^i$. A union bound over the sets of $\ell$ vertices and over the sets of $\ell+1$ pairs inside them gives
\[
 \Pr(\mathcal E)\le\sum_{\ell\le2p}\binom n\ell\binom{\binom\ell2}{\ell+1}\Bigl(\frac{2c}{n-1}\Bigr)^{\ell+1}
 \le\sum_{\ell\le2p}n^\ell\,\ell^{2\ell+2}\Bigl(\frac{2c}{n-1}\Bigr)^{\ell+1}
 \le\frac1n\exp\bigl(O_c(p\log(p+2))\bigr),
\]
which is $n^{-1+o(1)}$ because $p\log(p+2)=o(\log n)$.

\emph{The running time.} Outside $\mathcal E$ the simple cycles with fewer than $P$ vertices in the support are pairwise vertex-disjoint. This property depends on the support alone, and Theorem~\ref{thm:short-cycles} gives $O(n+m+N\sqrt L)=O_c(n\sqrt{\log n})$ time for every choice of directions, weights and source. On every input the time is $O(n+m+NL)=O_c(n\log n)$. The expected maximum running time is therefore at most $O_c(n\sqrt{\log n})+n^{-1+o(1)}\cdot O_c(n\log n)=O_c(n\sqrt{\log n})$.
\end{proof}

\subsection{The recursive search}\label{subsec:search}

The procedure $\Search$ is the bounded recursive search of~\cite[Algorithm~2]{GZ}, used with the scales \eqref{eq:scales} and with a new first step. In this subsection and in Sections~\ref{subsec:prepare}--\ref{subsec:time}, $G$ may be any simple directed graph with $N$ vertices, $M$ arcs, incoming and outgoing degrees at most $D$, nonnegative real weights and a source $s'$, for which the guards \eqref{eq:guards} hold. The search starts from the labels and the empty certificate lists of Algorithm~\ref{alg:driver}.

Walks from $s'$ are compared by length, then by number of arcs, called \emph{hops}, then lexicographically by the reversed sequence of their vertex identifiers. This is the \emph{witness order}. It is a total order on the walks from $s'$, because $G$ is simple; the walks from $s'$ to a reachable vertex have a least element. The \emph{signature} $\sig(\omega)$ of a walk $\omega$ consists of its length, its number of hops, its last vertex and its penultimate vertex (a fixed sentinel for the walk without arcs). Signatures are compared lexicographically, and the signature of an infinite label is larger than that of every walk. A signature lists the first four keys by which the witness order compares walks, so $\omega\le\omega'$ in the witness order implies $\sig(\omega)\le\sig(\omega')$. The algorithm stores the signature of each label and not the walk, which serves only in the proofs, and it compares labels, candidates and bounds through their signatures~\cite[Section~3.1]{GZ}. The \emph{candidate} along an arc $(u,v)$ with $\dhat(u)$ finite is the walk $\dhat(u)\cdot(u,v)$; its signature is computed from $\sig(\dhat(u))$ in constant time. An \emph{update} along $(u,v)$ is an attempt to replace $\dhat(v)$ by this candidate $c$. It is accepted exactly when
\begin{equation}\label{eq:guard}
 \sig(c)\le\sig(\dhat(v)),
\end{equation}
and labels change only by accepted updates. Equal signatures are accepted; the reason is given after (A4) below, and (W1) in Section~\ref{subsec:admissible} shows that an accepted update never makes a label larger in the witness order. Let $\pi(v)$ be the least walk from $s'$ to $v$ in the witness order, for every reachable vertex $v$. A vertex $v$ is \emph{complete} when $\dhat(v)=\pi(v)$. The goal of the search is to make every reachable vertex complete.

A \emph{bound} is a signature or $+\infty$. A walk is \emph{below} a bound $B$ if its signature is smaller than $B$, and a reachable vertex $v$ is below $B$ if $\pi(v)$ is. The vertices on $\pi(v)$, including $v$, are the \emph{ancestors} of $v$. A call $\Search(B,S,j)$ has a bound $B$, a level $j\ge0$ and a nonempty set $S$ of at most $A_j$ \emph{sources} whose labels are below $B$. Its \emph{targets} are the elements of
\[
 \calT(B,S)\coloneqq\{v:\ v\text{ is below }B\text{ and has an ancestor in }S\}.
\]
Every source is a target. The call is meant to complete the targets in increasing order of $\sig(\pi(v))$, up to a cut $\beta\le B$ that it returns. It requires two input conditions:
\begin{enumerate}[label=(I\arabic*)]
\item every target has a complete ancestor in $S$;
\item every vertex below $B$ that is not a target is complete.
\end{enumerate}
The top call $\Search(+\infty,\{s'\},e)$ satisfies (I1) and (I2)~\cite[Section~4.1]{GZ}, and its targets are the reachable vertices.

A call at level $0$ is a search with a priority queue, kept as a balanced search tree~\cite[Section~5.2]{GZ}. The queue initially holds the sources. The call repeatedly extracts the vertex of least signature, adds it to its output, performs the accepted updates with candidates below $B$ along its outgoing arcs and puts the updated heads into the queue, until the queue is empty or the output has $H_0$ vertices~\cite[Algorithm~2, lines 2--11]{GZ}. It returns as its cut the least key left in the queue, or $B$ if the queue is empty, together with its output and the remaining queue as its worklist. A call at a level $j\ge1$ is Algorithm~\ref{alg:search}. It first runs a \emph{preparation step}, which retires some of its sources and partitions the others into groups. It then repeatedly pulls at most $M_j$ entries from a worklist, runs a call of level $j-1$ on the pulled vertices, below the bound that separates them from the remaining entries, and performs the accepted updates along the arcs out of the vertices that this call has completed. It stops when the worklist is empty or its output has at least $H_j$ vertices.

A \emph{worklist} stores vertices with keys. The \emph{key} of a vertex is a copy of the signature that its label had when the vertex was inserted, and the label can later become smaller. A worklist holds at most one entry per vertex, and an insertion keeps the smaller key. A \emph{pull} removes the at most $M_j$ smallest entries and returns as its bound $z$ the least remaining key, or $B$ when none remain. Every key is below $B$, and keys of different vertices differ, because a key contains its vertex; so the removed keys are below $z$ and the remaining keys are at least $z$. We use the explicit worklists of~\cite[Appendix~B]{GZ}, in which an insertion has amortized cost $O(T)$: by~\cite[Proposition~B.1]{GZ} this cost is logarithmic in $(A_j+2DH_j)/M_j$, where $A_j+2DH_j$ is the bound on the number of records (stored entries) at level~$j$, and the last guard in \eqref{eq:guards} bounds this logarithm by $2T$~\cite[Section~5.2]{GZ}; Appendix~\ref{app:import}, items~2 to~4, shows that the record bound holds for this search.

The preparation step $\Prepare$ returns a set $Q\subseteq S$ of \emph{retired} sources, a set $W$ of vertices and a partition of $P_S\coloneqq S\setminus Q$ into \emph{groups}; at the end, the call adds to its output the vertices of $W$ whose labels are below its final cut (line~\ref{line:wnew}). A group is represented in the worklist by its \emph{representative}, a member whose label has the least signature when it is chosen~\cite[Section~4.1]{GZ}. When the representative is pulled, the other members with labels below the separating bound join the recursive call (line~\ref{line:expand}). When it is in the output of a recursive call and the group still has members, a new representative is chosen (line~\ref{line:reselect}). A retired source belongs to no group. The variable $\lambda$ is the lower cut of the remaining work, called $L$ in~\cite[Section~4.2]{GZ}. At the start of each iteration of the loop of line~\ref{line:loop} and after line~\ref{line:finalcut}, the targets below $\lambda$ that have an ancestor in $P_S$ are exactly the vertices in the outputs of the recursive calls made so far, these vertices are complete, and the worklist holds no key below $\lambda$~\cite[Appendix~D, (J1)--(J4)]{GZ}. A target without an ancestor in $P_S$ is complete by (A1) below, so every target below $\lambda$ is complete. An entry can remain for a vertex that a recursive call has completed; its label is then below $\lambda$, and line~\ref{line:filter} discards the entry. The call returns its final cut $\beta$, its output and its residual worklist. The return is \emph{full} if $\beta=B$ and \emph{partial} otherwise.

\begin{algorithm}[htbp]
\caption{$\Search(B,S,j)$}\label{alg:search}
\KwIn{bound $B$; sources $S$; level $j$; the graph $G$, the labels $\dhat$ and the certificate lists are global}
\KwOut{cut $\beta$; set of completed targets; residual worklist}
\If{$j=0$}{
  \Return the result of the priority-queue search from $S$ below $B$\Rem*{\cite[Algorithm~2, lines 2--11]{GZ}}\label{line:base}
}
$(Q,W,\mathit{groups})\gets\Prepare(B,S)$\Rem*{Algorithm~\ref{alg:prepare}; replaces~\cite[Algorithm~2, line 13]{GZ}}\label{line:prepare}
$U\gets\varnothing$;\quad $\calD\gets$ a worklist that holds the representative of each group\;\label{line:first}
$\lambda\gets$ the least key in $\calD$, or $B$ if $\calD=\varnothing$\;
\While{$\calD\ne\varnothing$ and $\size U<H_j$}{\label{line:loop}
  pull the at most $M_j$ smallest entries of $\calD$;\quad $z\gets$ their separating bound\;
  $S'\gets$ the pulled vertices whose labels are not below $\lambda$\;\label{line:filter}
  add to $S'$ the members with labels below $z$ of every group whose current representative was pulled\;\label{line:expand}
  \If{$S'\ne\varnothing$}{
    $(\beta',U',\calD')\gets\Search(z,S',j-1)$\;\label{line:child}
    merge $\calD'$ into $\calD$;\quad remove $U'$ from the groups\;
    \ForEach{arc $(u,v)$ with $u\in U'$ whose candidate $c$ has $z\le\sig(c)<B$}{
      \lIf{$\sig(c)\le\sig(\dhat(v))$}{$\dhat(v)\gets c$; insert $v$ into $\calD$}
    }
    insert into $\calD$ a new representative of every nonempty group whose representative is in $U'$\;\label{line:reselect}
    $U\gets U\cup U'$;\quad $\lambda\gets\beta'$\;
  }
}
\lIf{$\calD=\varnothing$}{$\lambda\gets B$}\label{line:finalcut}
insert into $\calD$ every source, retired or not, whose label is not below $\lambda$\;\label{line:restore}
$W_{\mathrm{new}}\gets\{v\in W\setminus U:\ \dhat(v)\text{ is below }\lambda\}$\;\label{line:wnew}
\ForEach{arc $(u,v)$ with $u\in W_{\mathrm{new}}$ whose candidate $c$ has $\lambda\le\sig(c)<B$}{
  \lIf{$\sig(c)\le\sig(\dhat(v))$}{$\dhat(v)\gets c$; insert $v$ into $\calD$}\label{line:wnewscan}
}
\Return $(\lambda,\,U\cup W_{\mathrm{new}},\,\calD)$\Rem*{the cut $\beta$ is the final $\lambda$}\label{line:return}
\end{algorithm}

Apart from the initialization of $U$, which~\cite{GZ} performs in line 1, lines~\ref{line:first}--\ref{line:return} are lines 14--31 of~\cite[Algorithm~2]{GZ}. There the procedure is $\mathcal R(B,S,j)$, the lower cut is $L$, the separating bound is $b$, the worklist is $\mathcal D_X$, and the sources, the cut, the output and the worklist of the recursive call are $S_c$, $\beta_c$, $U_c$ and $\mathcal D_c$.

Line~\ref{line:prepare} is the only change, and it creates two obligations. For correctness, the output of $\Prepare$ must have the properties that the proof of~\cite{GZ} uses. They are isolated next, and Section~\ref{subsec:admissible} proves them. For the running time, the work that depends on the preparation step must be charged: the step itself, the worklist entries of the groups in lines~\ref{line:first} and~\ref{line:reselect}, the expansions of the groups in line~\ref{line:expand}, and the computation and scan of $W_{\mathrm{new}}$ in lines~\ref{line:wnew}--\ref{line:wnewscan}. Section~\ref{subsec:time} charges it. All other costs are those of~\cite{GZ}.

In~\cite{GZ} the preparation step is an operation $\operatorname{Forest}(B,S)$ built from bounded shortest-path searches. The proof of~\cite[Theorem~4.1]{GZ} uses four properties of the output of this operation, which we isolate. The output of a preparation step in a call $\Search(B,S,j)$ is \emph{admissible} if the following hold when the step ends.
\begin{enumerate}[label=(A\arabic*)]
\item Every target has a complete ancestor in $P_S$, or is complete and belongs to $W$.
\item $W\subseteq\calT(B,S)$ and $\size W\le P\size Q$.
\item The groups partition $P_S$, and each group has fewer than $3P$ members.
\item The step changes labels only by accepted updates along arcs whose tails are targets and whose candidates are below $B$, and it creates no worklist entry.
\end{enumerate}
Property (A1) lets the search continue from $P_S$ alone, (A2) and (A3) limit the sizes of the sets that a call handles (Appendix~\ref{app:import}, items~2 and~3), and (A4) keeps the invariants of the calling search intact (Appendix~\ref{app:import}, item~4). Because the step creates no worklist entry, a vertex whose label it has set may still have to be inserted into a worklist by a later scan of the arcs out of a completed vertex, and the candidate of that scan can have the same signature as that label; this is why \eqref{eq:guard} accepts equal signatures~\cite[Section~3.1]{GZ}.

\begin{lemma}[Return contract]\label{lem:contract}
Suppose that the preparation step of every call at a level $j\ge1$ terminates with admissible output whenever the call satisfies \textup{(I1)} and \textup{(I2)}. Then every call $\Search(B,S,j)$ made during the top call satisfies \textup{(I1)} and \textup{(I2)}, terminates, and has the properties stated in~\cite[Theorem~4.1]{GZ}. In particular, its cut $\beta$ and its output $U$ satisfy:
\begin{enumerate}[label=\textup{(R\arabic*)}]
\item $U$ consists of the targets below $\beta$, and its vertices are complete;
\item if $\beta<B$ then $\size U\ge H_j\ge T\size S$, and if $\beta=B$ then $U=\calT(B,S)\supseteq S$;
\item the outputs of different calls of the same level are disjoint.
\end{enumerate}
\end{lemma}
\begin{proof}
Theorem~4.1 of~\cite{GZ} states these properties for the search whose preparation step is $\operatorname{Forest}$, under two hypotheses (F) and (F$'$) on its bounded searches~\cite[Section~4.1]{GZ}. Hypothesis (F) says that if a bounded search starts at a complete source and exhausts, that is, ends without being stopped, with fewer than $P$ retained vertices, then it completes every target that has this source as an ancestor and puts it in $W$; hypothesis (F$'$) describes the trees of these searches. The proof of that theorem~\cite[Appendix~D]{GZ} refers to the preparation step only through (A1)--(A4). Appendix~\ref{app:import} lists each place where the step enters that proof, with the property used there, and checks the other hypotheses of that theorem. It does the same for the two counting arguments of~\cite[Appendix~E]{GZ} that Section~\ref{subsec:time} uses, the bound~\cite[(53)]{GZ} and the colouring of~\cite[Appendix~E.2]{GZ}, and for the remaining work in the proof of~\cite[Lemma~5.1]{GZ}.
\end{proof}

\subsection{The preparation step}\label{subsec:prepare}

In~\cite{GZ} the searches of $\operatorname{Forest}$ are charged $G_p=\Theta(\sqrt{\log\log N})$ operations per retained vertex when $D$ is bounded~\cite[(36)]{GZ}, and the running time~\cite[(40)]{GZ} contains the factor $\sqrt{G_p}$. This is the factor $(\log\log n)^{1/4}$ in the sparse bound of~\cite[Theorem~1.1]{GZ}. In $\Prepare$ each source instead claims vertices by a depth-first search, at constant cost per claimed vertex, and its search stops when the source owns $P$ vertices or the depth-first search has ended.

The claimed vertices form connected components in the support of $G$. The first worklist entry of a group, created in line~\ref{line:first} of Algorithm~\ref{alg:search}, costs $O(T)$. In a component with at least $P$ vertices the groups are formed as in~\cite{GZ}, and this cost is $O(T/P)$ per claimed vertex. A component $C$ with fewer than $P$ vertices cannot pay in this way. $\Prepare$ therefore tries to retire one of its sources, because the analysis of~\cite{GZ} pays for the retired sources. Depth-first search does not visit vertices in order of distance, so a source cannot be retired merely because its search has ended. $\Prepare$ retires a source of $C$ in two cases, described after Algorithm~\ref{alg:prepare}, in which (A1) and (A2) remain true without it: another source of $C$ replaces it, or the targets that need it as a complete ancestor are themselves complete and are returned through $W$. If neither case occurs, $C$ contains two competing paths of claims that close a cycle with fewer than $P$ vertices (Lemma~\ref{lem:diamond}). $\Prepare$ then stores a certificate for this cycle at one of its arcs, unless the list of that arc already holds $b$ certificates. A stored certificate keeps its cycle from causing the same cost at that arc again, so the cost recurs at an arc only after $b$ different cycles with fewer than $P$ vertices have each caused it there (Lemma~\ref{lem:certificates}). This is how the short cycles of Theorem~\ref{thm:short-cycles} enter the running time (Theorem~\ref{thm:structural}).

For a directed path $\rho$ with at least one arc, let $\tau(\rho)$ be the triple formed by its length, its number of arcs and its penultimate vertex. Triples are ordered lexicographically. For a walk $\omega$ from $s'$ to the first vertex of $\rho$, the signature $\sig(\omega\cdot\rho)$ is obtained from $\sig(\omega)$, $\tau(\rho)$ and the last vertex of $\rho$ by two additions. If $\rho$ and $\rho'$ have the same first and last vertices, then
\begin{equation}\label{eq:tau}
 \sig(\omega\cdot\rho)<\sig(\omega\cdot\rho')\quad\text{if and only if}\quad\tau(\rho)<\tau(\rho'),
\end{equation}
and if $\omega'\le\omega$ in the witness order, then $\sig(\omega'\cdot\rho)\le\sig(\omega\cdot\rho)$.

Every arc has a list of at most $b$ certificates. A \emph{certificate} at an arc $(p,h)$ is a pair $(w,\tau(\rho))$, where $\rho$ is a directed path of $G$ from a vertex $w\ne h$ to $h$. It \emph{blocks} a candidate $c$ along $(p,h)$ if $\dhat(w)$ is finite and $\sig(\dhat(w)\cdot\rho)<\sig(c)$. In a call with bound $B$, an arc $(u,h)$ whose tail has a finite label passes the test $\Admit(u,h)$ if its candidate $c$ satisfies
\[
 c\text{ is below }B,\qquad \sig(c)\le\sig(\dhat(h)),\qquad\text{and no certificate at }(u,h)\text{ blocks }c .
\]
The test takes $O(1+b)$ time. It strengthens \eqref{eq:guard}. A blocked candidate is not the least walk to $h$, because the walk $\dhat(w)\cdot\rho$ has a smaller signature.

An execution of $\Prepare$ keeps the following records. The search of a source $r$ \emph{claims} a vertex $h$ by setting its \emph{owner} $\owner(h)$ to $r$ (line~\ref{line:claim} of Algorithm~\ref{alg:prepare}). Each source owns itself before the searches start and no search claims it. A non-source becomes claimed when it gets an owner; its \emph{discovery arc} $\disc(h)$ is the arc along which it was claimed. Set $\disc(h)=\bot$ for a source, where $\bot$ is a marker distinct from every arc. The discovery arcs of the vertices owned by a source $r$ form a tree rooted at $r$, the \emph{first-claim tree} of $r$. An arc is \emph{scanned} when a search examines it at its tail; this happens at most once in an execution.

\begin{algorithm}[htbp]
\caption{$\Prepare(B,S)$}\label{alg:prepare}
\KwIn{bound $B$; sources $S$ in a fixed order; the graph $G$, the labels $\dhat$, the certificate lists, $P$ and $b$ are global}
\KwOut{retired sources $Q$; vertex set $W$; a partition of $S\setminus Q$ into groups}
$Q\gets\varnothing$;\quad $W\gets\varnothing$;\quad $\mathit{groups}\gets\varnothing$;\quad $\mathit{certs}\gets\varnothing$\;
\lForEach{$r\in S$}{$\owner(r)\gets r$; $\disc(r)\gets\bot$\Rem*[f]{each source owns itself}}
\ForEach(\Rem*[f]{first claims}){$r\in S$ in order}{\label{line:search-loop}
  $\mathit{stack}\gets(r)$;\quad $\mathit{count}\gets1$\;
  \While{$\mathit{stack}$ is nonempty and $\mathit{count}<P$}{\label{line:stop}
    $u\gets$ the top of $\mathit{stack}$\;
    \eIf{$u$ has an outgoing arc that has not been scanned}{
      $(u,h)\gets$ the next such arc\;
      \If{$h$ is unclaimed and $\Admit(u,h)$}{\label{line:test}
        $\dhat(h)\gets\dhat(u)\cdot(u,h)$;\quad $\owner(h)\gets r$;\quad $\disc(h)\gets(u,h)$\;\label{line:claim}
        push $h$ on $\mathit{stack}$;\quad $\mathit{count}\gets\mathit{count}+1$\;
      }
    }{
      pop $u$ from $\mathit{stack}$\;
    }
  }
}
\ForEach{connected component $C$ of the claimed vertices}{\label{line:components}
  \eIf(\Rem*[f]{large component}){$\size C\ge P$}{
    partition the edges of a spanning tree of $C$ into connected pieces with at least $P$ and fewer than $3P$ vertices\;\label{line:pieces}
    assign each source of $C$ to one piece that contains it\;
    add to $\mathit{groups}$ each nonempty set of sources assigned to a piece\;\label{line:largegroup}
  }(\Rem*[f]{small component}){
    $\mathit{Dev}\gets\{(t,h):\ t,h\in C,\ (t,h)\ne\disc(h),\ \Admit(t,h)\}$\;\label{line:dev}
    $\mathit{Deviating}\gets\{\owner(t):\ (t,h)\in\mathit{Dev}\}$\;\label{line:mark}
    \uIf{some arc of $\mathit{Dev}$ enters a source}{\label{line:ifsource}
      $Q\gets Q\cup\{h\}$ for one arc $(t,h)\in\mathit{Dev}$ with $h\in S$;\quad $W\gets W\cup C$\;\label{line:retire-head}
    }
    \uElseIf{some source of $C$ is not in $\mathit{Deviating}$}{
      $Q\gets Q\cup\bigl((C\cap S)\setminus\mathit{Deviating}\bigr)$;\quad $W\gets W\cup C$\;\label{line:retire-unmarked}
    }
    \Else{
      $\mathit{certs}\gets\mathit{certs}\cup\Learn(C,\mathit{Dev})$\Rem*{Algorithm~\ref{alg:learn}}\label{line:learn}
    }
    \lIf{$(C\cap S)\setminus Q\ne\varnothing$}{add $(C\cap S)\setminus Q$ to $\mathit{groups}$}\label{line:smallgroup}
  }
}
append every certificate in $\mathit{certs}$ to the list of its arc\;\label{line:install}
\Return $(Q,W,\mathit{groups})$\;
\end{algorithm}

The components of line~\ref{line:components} are those of the claimed vertices in the support of $G$, and line~\ref{line:pieces} uses the partition of~\cite[Lemma~B.2]{GZ}. A search that leaves the loop of line~\ref{line:stop} with an empty stack has scanned every arc out of every vertex that its source owns. Labels change only in line~\ref{line:claim} and certificate lists only in line~\ref{line:install}. Between the end of the loop of line~\ref{line:search-loop} and line~\ref{line:install} the outcome of every test $\Admit$ is therefore fixed, and we say that an arc passes or fails the test \emph{after the first claims}. The arcs in $\mathit{Dev}$ are the \emph{deviations} of $C$: the arcs inside $C$, other than discovery arcs, that pass the test after the first claims. A source has no discovery arc, so every arc inside $C$ that enters a source and passes the test is a deviation. A source that owns the tail of a deviation is \emph{deviating}.

The three cases for a small component $C$ have the following roles. Only complete sources matter for (A1), because (I1) gives every target a complete ancestor in $S$; $\Prepare$ cannot test completeness, and retiring a source that is not complete takes no such ancestor from any target. Lemma~\ref{lem:admissible} proves the following two statements about complete sources. If a deviation $(t,h)$ enters a complete source $h$, then the owner of $t$ is a complete ancestor of $h$. This owner differs from $h$ (proof of Lemma~\ref{lem:admissible}) and lies in $C$, because its first-claim tree contains $t$. As line~\ref{line:retire-head} retires only $h$ in $C$, the owner replaces $h$ as a complete ancestor in $P_S$, and $h$ is retired. Line~\ref{line:retire-head} also puts $C$ into $W$; (A1) does not need this, and it is done only to treat the two retirement cases alike. If a complete source $r$ of $C$ is not deviating, then every target that has $r$ as an ancestor lies in the first-claim tree of $r$ and is complete. These targets belong to $C\subseteq W$, as (A1) requires, and $r$ is retired. In both cases $C$ contains a retired source, to which the analysis charges the first worklist entry of the group of $C$. In the remaining case $\Learn$ finds a diamond in $C$, as defined next, and returns the certificate of its cycle unless the list of its losing arc is full.

A small component in which no source is retired contains two paths of a first-claim tree, one of them extended by a deviation, in the following configuration (Lemma~\ref{lem:diamond}).
\begin{definition}\label{def:diamond}
A \emph{diamond} is a pair $(\Omega,\Lambda)$ of internally vertex-disjoint directed paths of $G$ from a vertex $u$ to a vertex $h\ne u$ with $\tau(\Omega)<\tau(\Lambda)$. Its \emph{fork} is $u$, its \emph{arms} are $\Omega$ and $\Lambda$, and its \emph{losing arc} is the last arc of $\Lambda$. Its \emph{cycle} is the union of the two arms in the support of $G$.
\end{definition}
The arms differ and $G$ is simple, so the cycle of a diamond is a simple cycle with at least three vertices. A simple cycle $K$ \emph{competes at} an arc $f$ if some diamond has cycle $K$ and losing arc $f$. Let $K$ compete at $f=(p,h)$. For a vertex $w\ne h$ of $K$, let $\Lambda_w$ be the path from $w$ to $h$ along $K$ that ends with $f$, and $\Omega_w$ the other path from $w$ to $h$ along $K$. The vertex $w$ is a \emph{fork of} $(K,f)$ if both are directed paths of $G$ and $\tau(\Omega_w)<\tau(\Lambda_w)$. These are the forks of the diamonds with cycle $K$ and losing arc $f$. Let $w^*$ be the fork $w$ of $(K,f)$ for which $\Lambda_w$ has the fewest arcs. The \emph{certificate of} $(K,f)$ is the pair $(w^*,\tau(\Omega_{w^*}))$, a certificate at $f$. With this choice of $w^*$, the single certificate prevents every later execution of $\Learn$ from finding a diamond with cycle $K$ and losing arc $f$ (Lemma~\ref{lem:certificates}(a)). Two traversals of $K$ that start at $h$ and accumulate lengths and hop counts compute it in $O(D\size{V(K)})$ time.

\begin{algorithm}[htbp]
\caption{$\Learn(C,\mathit{Dev})$}\label{alg:learn}
\KwIn{a small component $C$ in which every source is deviating and no deviation enters a source; its set $\mathit{Dev}$ of deviations; the owners and the discovery arcs are those of the calling execution of $\Prepare$}
\KwOut{a set with at most one certificate}
$r\gets$ the source of $C$ whose search ran first in line~\ref{line:search-loop} of $\Prepare$\;
$(t,h)\gets$ an arc of $\mathit{Dev}$ with $\owner(t)=r$\Rem*{exists because $r$ is deviating}
$u\gets$ the lowest common ancestor of $t$ and $h$ in the first-claim tree of $r$\Rem*{Lemma~\ref{lem:diamond}}
$\Lambda\gets$ the tree path from $u$ to $h$;\quad $\Omega\gets$ the tree path from $u$ to $t$, followed by $(t,h)$\;
$f\gets\disc(h)$;\quad $K\gets$ the cycle formed by $\Omega$ and $\Lambda$\;
\lIf{the list of $f$ holds $b$ certificates}{\Return $\varnothing$}\label{line:full}
\Return $\{$the certificate of $(K,f)\}$\;\label{line:record}
\end{algorithm}

\subsection{Correctness}\label{subsec:admissible}

By Lemma~\ref{lem:contract}, the search is correct once the output of $\Prepare$ is admissible. The proofs use three facts about labels from~\cite[Lemma~3.1 and Appendix~A]{GZ}, which hold because labels change only by accepted updates.
\begin{enumerate}[label=(W\arabic*)]
\item If the candidate $c$ along an arc into $v$ satisfies \eqref{eq:guard}, then $c\le\dhat(v)$ in the witness order. Hence the label of a vertex never increases in the witness order.
\item The walk $\pi(v)$ is a simple path, and each of its prefixes is $\pi(y)$ for its last vertex $y$.
\item A complete vertex remains complete, and every vertex on $\pi(v)$ is complete when $v$ is.
\end{enumerate}
Every label and every candidate at $v$ is a walk from $s'$ to $v$. It is therefore at least $\pi(v)$ in the witness order, and its signature is at least $\sig(\pi(v))$.

\begin{lemma}\label{lem:frozen}
Consider one execution of $\Prepare(B,S)$ in a call that satisfies \textup{(I1)} and \textup{(I2)}.
\begin{enumerate}[label=\textup{(\alph*)}]
\item Each label change in line~\ref{line:claim} is an accepted update along an arc whose tail is a target and whose candidate is below $B$. Every claimed vertex is a target.
\item From the moment a vertex $h$ is claimed with discovery arc $(u,h)$ until the end of the execution, $\dhat(h)=\dhat(u)\cdot(u,h)$. Hence the label of a vertex in the first-claim tree of a source $r$ is $\dhat(r)$ followed by its tree path from $r$.
\item Let $u$ be a claimed vertex and $(u,h)$ an arc. The candidate along $(u,h)$ is the same at all times after $u$ is claimed. If a certificate blocks it at some time, the certificate blocks it until the end of the execution.
\item If $u$ lies in a small component, $(u,h)$ is not the discovery arc of $h$, and $h$ was unclaimed when the search of the owner of $u$ scanned $(u,h)$, then $(u,h)$ fails the test after the first claims. This holds in particular when $h$ is unclaimed after the first claims.
\item No certificate blocks a candidate equal to $\pi(h)$.
\end{enumerate}
\end{lemma}
\begin{proof}
(a) Line~\ref{line:claim} applies an update that passed $\Admit$, so the update is accepted and its candidate $c$ is below $B$. The sources are targets. It remains to show that $h$ is a target when the tail $u$ is one; this is the containment argument of~\cite[Lemma~4.2]{GZ}. The vertex $h$ is below $B$ because $\sig(\pi(h))\le\sig(c)$. If $h$ were not a target, it would be complete by (I2), so $c\le\pi(h)$ by (W1) and therefore $c=\pi(h)$. By (W2) the prefix $\dhat(u)$ of $c$ would equal $\pi(u)$, and an ancestor of $u$ in $S$ would be an ancestor of $h$, which makes $h$ a target.

(b) After $h$ is claimed, no search claims it again, so its label does not change. The same holds for $u$, which was claimed earlier.

(c) The first assertion follows from (b). For a certificate $(w,\tau(\rho))$, the label $\dhat(w)$ does not increase in the witness order by (W1), so $\sig(\dhat(w)\cdot\rho)$ does not increase, and no certificate is removed from a list.

(d) The owner of $u$ owns fewer than $P$ vertices, since all of them lie in the component of $u$. Its search therefore ended with an empty stack, so it scanned $(u,h)$ at a time when $u$ was claimed. The vertex $h$ was unclaimed then. If the arc had passed the test at that time, the search would have claimed $h$ along $(u,h)$, and $(u,h)$ would be the discovery arc of $h$. So the arc failed the test at that time. By (c) the candidate is the same after the first claims and a certificate that blocked it still blocks it, and by (W1) the label of $h$ has not increased. So the arc still fails the test after the first claims. A vertex that is unclaimed after the first claims has no discovery arc and was unclaimed at every earlier time.

(e) If a certificate $(w,\tau(\rho))$ at an arc into $h$ blocks $c$, then $\dhat(w)\cdot\rho$ is a walk from $s'$ to $h$ with a smaller signature than $c$, so $c\ne\pi(h)$.
\end{proof}

The next lemma shows that $\Learn$ is well defined and finds a diamond.

\begin{lemma}\label{lem:diamond}
Consider an execution of $\Learn(C,\mathit{Dev})$ in the setting of Lemma~\ref{lem:frozen}, with $r,(t,h),\Omega,\Lambda,f,K$ as in Algorithm~\ref{alg:learn}. The vertex $h$ is not a source, is owned by $r$, and is not an ancestor of $t$ in the first-claim tree of $r$. The pair $(\Omega,\Lambda)$ is a diamond whose cycle $K$ has fewer than $P$ vertices, all in $C$, and its losing arc is $f$.
\end{lemma}
\begin{proof}
No deviation of $C$ enters a source, so $h$ is not a source. Let $c$ be the candidate along $(t,h)$.

Suppose that $h$ is owned by a source $r'\ne r$. Then $r'$ lies in $C$, so the search of $r'$ ran after the search of $r$ by the choice of $r$, and $h$ was unclaimed when the search of $r$ scanned $(t,h)$, which is not a discovery arc. By Lemma~\ref{lem:frozen}(d) the arc $(t,h)$ would fail the test after the first claims, but it is a deviation. So $r$ owns $h$.

If $h$ were an ancestor of $t$ in the first-claim tree of $r$, Lemma~\ref{lem:frozen}(b) would give $c=\dhat(h)\cdot\rho\cdot(t,h)$ for the tree path $\rho$ from $h$ to $t$. Then $c$ would have more hops than $\dhat(h)$ and at least its length, contrary to \eqref{eq:guard}.

Thus $u\ne h$. The tree paths from $u$ to $t$ and from $u$ to $h$ share only $u$, and $h$ is not on the first of them, so $\Omega$ and $\Lambda$ are internally vertex-disjoint. The last arc of $\Lambda$ is the discovery arc $f=(p,h)$ of $h$, and $t\ne p$ because $(t,h)$ is not a discovery arc and $G$ is simple. By Lemma~\ref{lem:frozen}(b), $\dhat(h)=\dhat(u)\cdot\Lambda$ and $c=\dhat(u)\cdot\Omega$. These walks have different penultimate vertices, so \eqref{eq:guard} gives $\sig(\dhat(u)\cdot\Omega)<\sig(\dhat(u)\cdot\Lambda)$, and $\tau(\Omega)<\tau(\Lambda)$ follows from \eqref{eq:tau}. All vertices of the two arms are owned by $r$ and lie in $C$.
\end{proof}

\begin{lemma}\label{lem:admissible}
In a call that satisfies \textup{(I1)} and \textup{(I2)}, $\Prepare$ terminates with admissible output.
\end{lemma}
\begin{proof}
Every loop of $\Prepare$ is finite, and $\Learn$ is well defined by Lemma~\ref{lem:diamond}. Property (A4) is Lemma~\ref{lem:frozen}(a), together with the facts that labels change only in line~\ref{line:claim} and that $\Prepare$ creates no worklist entry.

(A3). Every source lies in exactly one component. A source of a large component is assigned to one piece, a piece has fewer than $3P$ vertices, and line~\ref{line:largegroup} adds only nonempty sets. The sources of a small component outside $Q$ form one group with fewer than $P$ members.

(A2). The set $W$ consists of claimed vertices, which are targets by Lemma~\ref{lem:frozen}(a). It is a union of small components, each of which has fewer than $P$ vertices and contains a retired source, so $\size W\le P\size Q$.

(A1). Let $v$ be a target. By (I1) it has a complete ancestor $r\in S$. If $r\notin Q$ we are done. Otherwise $r$ was retired in line~\ref{line:retire-unmarked} or in line~\ref{line:retire-head}, in a small component $C$.

\emph{Line~\ref{line:retire-unmarked}.} Here $r$ is not deviating. Write $\pi(v)$ as $\pi(r)$ followed by a path $x_0x_1\cdots x_\ell$ with $x_0=r$ and $x_\ell=v$. We show by induction on $i$ that $x_i$ lies in the first-claim tree of $r$ and is complete. This holds for $i=0$. Let $i<\ell$. By (W2) the candidate along $(x_i,x_{i+1})$ is $\pi(x_i)\cdot(x_i,x_{i+1})=\pi(x_{i+1})$. It is below $B$ because $v$ is, it satisfies \eqref{eq:guard} because every label of $x_{i+1}$ is at least $\pi(x_{i+1})$, and no certificate blocks it by Lemma~\ref{lem:frozen}(e). So the arc passes the test after the first claims, and $x_{i+1}$ is claimed by Lemma~\ref{lem:frozen}(d). Then $x_{i+1}\in C$. If $(x_i,x_{i+1})$ were not the discovery arc of $x_{i+1}$, it would be a deviation whose tail is owned by $r$, and $r$ would be deviating. Hence it is the discovery arc of $x_{i+1}$. So $x_{i+1}$ is owned by $r$, and $\dhat(x_{i+1})=\pi(x_{i+1})$ by Lemma~\ref{lem:frozen}(b). For $i=\ell$ we obtain that $v$ is complete and belongs to $C\subseteq W$.

\emph{Line~\ref{line:retire-head}.} Here $r$ is the head of a deviation $(t,r)$. Let $r'$ be the owner of $t$. By Lemma~\ref{lem:frozen}(b) the candidate along $(t,r)$ is $c=\dhat(r')\cdot\rho\cdot(t,r)$, where $\rho$ is the tree path from $r'$ to $t$. If $r'=r$, then $c$ would have more hops than $\dhat(r)$ and at least its length, contrary to \eqref{eq:guard}. So $r'\ne r$. Since $r$ is complete, (W1) gives $c\le\pi(r)$ and therefore $c=\pi(r)$. By (W2) the prefix $\dhat(r')$ of $\pi(r)$ equals $\pi(r')$. Thus $r'$ is a complete ancestor of $r$, hence of $v$. Line~\ref{line:retire-head} retires only $r$ in $C$, so $r'\in P_S$.
\end{proof}

\begin{corollary}\label{cor:exact}
If the guards hold, the call $\Search(+\infty,\{s'\},e)$ makes every reachable vertex of $G$ complete.
\end{corollary}
\begin{proof}
By Lemmas~\ref{lem:admissible} and~\ref{lem:contract} the top call satisfies (R1) and (R2). A partial return would have $\size U\ge H_e=T^32^{eT}>N$. So the top call returns fully, and its output $\calT(+\infty,\{s'\})$ consists of all reachable vertices, which are complete.
\end{proof}

\subsection{Running time}\label{subsec:time}

The cost of the search is that of~\cite{GZ} except for the work listed after Algorithm~\ref{alg:search}: the preparation steps, the worklist entries and the expansions of the groups, and the computation and scans of $W_{\mathrm{new}}$. The first worklist entry of a group in a small component needs a new charge, which Lemma~\ref{lem:diamond} and the following lemma supply.

\begin{lemma}\label{lem:certificates}
\begin{enumerate}[label=\textup{(\alph*)}]
\item Once the certificate of a pair $(K,f)$ is in the list of $f$, no execution of\/ $\Learn$ finds a diamond with cycle $K$ and losing arc $f$.
\item The certificates in the list of an arc $f$ belong to pairs $(K,f)$ with distinct cycles $K$, each with fewer than $P$ vertices and competing at $f$.
\item If an execution of\/ $\Learn$ returns in line~\ref{line:full}, then more than $b$ simple cycles with fewer than $P$ vertices compete at its arc $f$.
\end{enumerate}
\end{lemma}
\begin{proof}
(a) Suppose that a later execution of $\Prepare$ runs $\Learn$ and finds a diamond $(\Omega,\Lambda)$ with cycle $K$ and losing arc $f=(p,h)$. Its fork $u'$ is a fork of $(K,f)$, with $\Lambda=\Lambda_{u'}$. The path $\Lambda_{w^*}$ has at most as many arcs as $\Lambda_{u'}$, and both run along $K$ and end with $f$, so $\Lambda_{w^*}$ is a suffix of $\Lambda$. By Lemma~\ref{lem:diamond}, $\Lambda$ is a path of a first-claim tree and $f$ is the discovery arc of $h$. By Lemma~\ref{lem:frozen}(b), the candidate along $f$ when $h$ was claimed was $c=\dhat(w^*)\cdot\Lambda_{w^*}$, for the label that $w^*$ had at that time. The certificate $(w^*,\tau(\Omega_{w^*}))$ was in the list of $f$ during that execution. Since $\tau(\Omega_{w^*})<\tau(\Lambda_{w^*})$, \eqref{eq:tau} gives $\sig(\dhat(w^*)\cdot\Omega_{w^*})<\sig(c)$. The certificate blocked $c$, so $h$ was not claimed along $f$, a contradiction.

(b) Every certificate in a list is the certificate of a pair $(K,f)$ found by $\Learn$, and Lemma~\ref{lem:diamond} shows that $K$ has fewer than $P$ vertices and competes at $f$. Two executions of $\Learn$ within one execution of $\Prepare$ act on disjoint components, hence on different arcs. A later execution finds a different cycle by (a).

(c) By (a) the cycle of the diamond found by this execution differs from the $b$ cycles of (b). It has fewer than $P$ vertices and competes at $f$ by Lemma~\ref{lem:diamond}.
\end{proof}

Call an arc $f$ \emph{$b$-heavy} if more than $b$ simple cycles with fewer than $P$ vertices compete at $f$. Let $\nu_b$ be the largest number of pairwise vertex-disjoint simple cycles, each with fewer than $P$ vertices and competing at some $b$-heavy arc. The number $\nu_b$ depends on $G$, its weights and $b$, and only the analysis uses it. A cycle that competes at an arc contains the edge of that arc. So if every edge of the support of $G$ lies on at most $b$ simple cycles with fewer than $P$ vertices, then no arc is $b$-heavy and $\nu_b=0$.

\begin{theorem}\label{thm:structural}
Fix $D$ and $b$, and let $G$ be a simple directed graph with nonnegative real weights, a source $s'$ and incoming and outgoing degrees at most $D$, as in Section~\ref{subsec:search}. If the guards hold, $\Search(+\infty,\{s'\},e)$ on $G$ takes $O_{D,b}\bigl(N\sqrt L+\nu_bL\bigr)$ time.
\end{theorem}
\begin{proof}
For a call $X$ at a level $j\ge1$, write $S_X$, $Q_X$, $W_X$ and $U_X$ for its sources, its retired sources, its set $W$ and its output, and $V_X$ for the set of vertices claimed by its preparation step. Then $\size{V_X}\le P\size{S_X}$, and $V_X$ consists of targets by Lemma~\ref{lem:frozen}(a), so $V_X\subseteq U_X$ when $X$ returns fully, by (R2); when $X$ returns partially, (R2) gives $\size{U_X}\ge T\size{S_X}$. There are $e+1$ levels, and $e+1\le L/T+2$. Table~\ref{tab:charges} lists the charges; we justify the rows in order.

\begin{table}[ht]
\centering\small
\begin{tabular}{@{}p{0.34\linewidth}p{0.30\linewidth}p{0.28\linewidth}@{}}
\toprule
Work & Bound & Reason \\
\midrule
executions of $\Prepare$ and $\Learn$ & $O_{D,b}(N(e+1))$ & (R2), (R3) \\
first entry of a group: & & \\
\quad partial return & $O(N(e+1))$ & (R2) \\
\quad large component & $O(N(e+1)T/P)$ & pieces have at least $P$ vertices \\
\quad small component, retired source & $O(T(N+M)+Ne)$ & \cite[(53)]{GZ} \\
\quad small component, $\Learn$ & $O(T(bM+(e+1)\nu_b))$ & Lemmas~\ref{lem:diamond} and~\ref{lem:certificates} \\
scans of the sets $W_{\mathrm{new}}$ & $O(DP(N+M+Ne/T))$ & (A2), \cite[(53)]{GZ} \\
later entries of groups & $O(T)$ per group, plus $O(MT)$ & \cite[Appendix~E.2]{GZ} \\
expansions of groups in line~\ref{line:expand} & $O(N(e+1))$; $O(P)$ per entry & (R2), \cite[Appendix~E.2]{GZ} \\
all other work of $\Search$ & the bound \eqref{eq:inherited} & \cite[Lemma~5.1]{GZ} \\
\bottomrule
\end{tabular}
\caption{The charges in the proof of Theorem~\ref{thm:structural}.}\label{tab:charges}
\end{table}

\emph{Executions of $\Prepare$.} By (R3) the sets $V_X$ of the fully returning calls of one level are disjoint. A partially returning call has $\size{V_X}\le P\size{S_X}\le\size{U_X}$ by (R2). Hence $\sum_X\size{V_X}\le2N(e+1)$. The searches scan each arc out of a claimed vertex once and evaluate one test per arc. The components, the spanning trees, the pieces~\cite[Lemma~B.2]{GZ}, the sets $\mathit{Dev}$ and the executions of $\Learn$, which read the tree paths from the discovery arcs, take time linear in the number of claimed vertices and their arcs, with a factor $O(1+b)$ for the tests.

\emph{First entries.} Line~\ref{line:first} of Algorithm~\ref{alg:search} selects the representative of each group and inserts it, at cost $O(P+T)=O(T)$. We count the groups.
\begin{itemize}
\item A partially returning call has at most $\size{S_X}\le\size{U_X}/T$ groups.
\item The pieces of a large component are edge-disjoint and have at least $P-1$ edges each, so a fully returning call has at most $\size{V_X}/(P-1)$ groups in large components.
\item A small component that retires a source is charged to that source. The bound
\[
 \sum_X\size{Q_X}\le N+M+Ne/T
\]
holds here. It is~\cite[(53)]{GZ} with the parameters used in the proof of~\cite[Lemma~5.1]{GZ}, and its proof uses (A3), (A4), (R2), (R3) and the fact that every arc causes at most one insertion from a completed tail (Appendix~\ref{app:import}, item~6).
\item An execution of $\Learn$ on a small component either yields a certificate or returns in line~\ref{line:full}. At most $bM$ certificates are stored during the whole computation. In the second case the component contains a simple cycle with fewer than $P$ vertices that competes at a $b$-heavy arc, by Lemmas~\ref{lem:diamond} and~\ref{lem:certificates}(c). Such components in fully returning calls of one level are pairwise disjoint, so each level has at most $\nu_b$ of them.
\end{itemize}

\emph{The sets $W_{\mathrm{new}}$.} Line~\ref{line:wnew} of Algorithm~\ref{alg:search} and the loop after it scan $O(D\size{W_X})$ arcs, and $\size{W_X}\le P\size{Q_X}$ by (A2).

\emph{Later entries.} Line~\ref{line:reselect} of Algorithm~\ref{alg:search} inserts a new representative of a group after its previous representative was returned in $U'$, at cost $O(P+T)$. A partially returning call does this at most $\size{S_X}$ times. For fully returning calls,~\cite[Appendix~E.2]{GZ} charges one such insertion to each group and every further one to an edge of the support of $G$, and charges each edge at most once in the whole computation. That argument requires the groups of a call to lie in pairwise edge-disjoint connected subgraphs of the support of $G$ whose vertices are targets of the call (Appendix~\ref{app:import}, item~7). Here these subgraphs are the pieces of the large components and the spanning trees of the small ones.

\emph{Expansions of groups.} Line~\ref{line:expand} of Algorithm~\ref{alg:search} expands a group when its representative is pulled: it goes through the members of the group, at cost $O(P)$ by (A3). The representative, which is in the output of no earlier recursive call, passes the filter of line~\ref{line:filter} by the invariant of $\lambda$ in Section~\ref{subsec:search} and becomes a source of the recursive call of line~\ref{line:child}. If this call returns partially, its output $U'$ has at least $T\size{S'}$ vertices by (R2), and, since $P\le T$, these vertices pay for the expansions of the at most $\size{S'}$ groups whose representatives are in $S'$; by (R3) this costs $O(N(e+1))$ in total. If it returns fully, it returns the representative, and the expansion is charged to the insertion of the next representative of the group or, if the group has become empty, to its first entry~\cite[Appendix~E.2]{GZ}; each worklist entry of a group receives at most one such charge of $O(P)$.

\emph{All other work.} The remaining work is that of Algorithm~\ref{alg:search}, apart from the work counted above, together with the array setup, the per-level arrays and the one-time sorting of the outgoing lists. The proof of~\cite[Lemma~5.1]{GZ} bounds it by
\begin{equation}\label{eq:inherited}
 O\bigl(N+M\log(2+D)+N(e+1)(D+1)+MT+ND\log(2+TD)\bigr),
\end{equation}
using the preparation step only through (A4), the bound on the sources of the recursive calls (Appendix~\ref{app:import}, item~3) and the properties of~\cite[Theorem~4.1]{GZ} that Lemma~\ref{lem:contract} provides (Appendix~\ref{app:import}, item~8). The bound \eqref{eq:inherited} is~\cite[(17)]{GZ} without its terms that contain the cost $h_{\mathrm{loc}}$ per retained vertex of a search or the factor $T/P$; in~\cite{GZ} they pay for the bounded searches of $\operatorname{Forest}$, the initial representatives and the reselections within pieces, which the rows above replace. The remaining terms pay, in the order of \eqref{eq:inherited}, for the array setup; for the one-time sorting of the outgoing lists~\cite[Section~3]{GZ}, which the search inherits and on which $\Prepare$ does not depend; for the ordinary scans and merges, the per-level arrays of~\cite[Appendix~F.1]{GZ} and the restorations of sources in line~\ref{line:restore}; for the insertions from completed tails; and for the base search trees (proof of~\cite[Lemma~5.1]{GZ}).

\emph{Total.} For fixed $D$ and $b$ we have $M\le DN$ and $T\le1024(P+1)$, so the sum of all rows is $O(N(e+1)+NT+T(e+1)\nu_b)$. Since $T\ge\sqrt L$, we get $N(e+1)\le N\sqrt L+2N$ and $T(e+1)\le L+2T$. Since $T\le\sqrt L+4096$, the total is $O(N\sqrt L+\nu_bL)$.
\end{proof}

The parameter arithmetic, the allocation of arrays on demand and the word sizes are those of~\cite[Appendix~F]{GZ}. The certificate lists add $O(bM)$ words, and the records of an execution of $\Prepare$ are initialized only at the vertices that it claims: each execution uses a fresh token, as in~\cite[Appendix~F.1]{GZ}, and a record with an older token reads as unclaimed.

\subsection{Short cycles in the routing graph}\label{subsec:cycles}

It remains to prove Theorem~\ref{thm:short-cycles}. Let $G$ be the routing graph of $H$ from Section~\ref{subsec:driver}. The hypothesis of the theorem concerns the support of $H$, while $\nu_b$ is defined on $G$; the next lemma relates the cycles of the two graphs.

\begin{lemma}\label{lem:routing}
Contracting every path of ports maps each simple cycle $\Gamma$ of the support of $G$ to a closed walk in the support of $H$ that has at most $\size{V(\Gamma)}$ edges and uses no edge twice. Different simple cycles of the support of $G$ are mapped to different closed walks.
\end{lemma}
\begin{proof}
A path of ports contains no cycle, so $\Gamma$ has edges that join ports of different vertices of $H$. Each of them corresponds to an edge of the support of $H$, and different edges of $\Gamma$ correspond to different support edges, because a support edge has only one pair of ports. This gives the closed walk. Conversely, two consecutive edges of the closed walk at a vertex $x$ determine two ports of $x$, and the path of ports of $x$ contains exactly one path between them. So the closed walk determines $\Gamma$.
\end{proof}

\begin{proof}[Proof of Theorem~\ref{thm:short-cycles}]
\emph{The guards.} Let $D=3$. If $L<4096$, then $T=4096>L$ and a guard fails. If $L\ge4096$, all guards hold: $T\le L$; $T^2\le\max\{2^{24},4L\}\le N$; $P\ge4$; $16(3P+1)\le3T/64+16\le T$; and the last guard divided by $M_j$ reads $2+(T/16)(1+2DT)2^T\le2^{2T}$. So if a guard fails, then $N$ is bounded and line~\ref{line:fallback} of Algorithm~\ref{alg:driver} takes $O(n+m)$ time.

\emph{All inputs.} If the guards hold, Corollary~\ref{cor:exact} and the routing graph give the correct distances, and Theorem~\ref{thm:structural} with $D=3$ bounds the time of line~\ref{line:topcall} of Algorithm~\ref{alg:driver}. All other lines take $O(n+m)$ time. Disjoint cycles have at least three vertices each, so $\nu_b\le N/3$, and the running time is $O_b(n+m+NL)$ on every input.

\emph{Disjoint short cycles.} Now suppose that the simple cycles with fewer than $P$ vertices in the support of $H$ are pairwise vertex-disjoint. Let $\Gamma$ be a simple cycle with fewer than $P$ vertices in the support of $G$. By Lemma~\ref{lem:routing} its closed walk in the support of $H$ uses no edge twice, so its edges decompose into simple cycles, each with fewer than $P$ vertices. The closed walk is connected, so if there were more than one such cycle, two of them would share a vertex. Hence the closed walk is a single simple cycle $\overline\Gamma$. By Lemma~\ref{lem:routing}, different cycles $\Gamma$ and $\Gamma'$ give different cycles $\overline\Gamma$ and $\overline{\Gamma'}$. These are vertex-disjoint, so $\Gamma$ and $\Gamma'$ pass through ports of different vertices of $H$ and share no edge. Thus every edge of the support of $G$ lies on at most one simple cycle with fewer than $P$ vertices. No arc of $G$ is $b$-heavy, $\nu_b=0$, and the running time is $O_b(n+m+N\sqrt L)$.
\end{proof}

Theorem~\ref{thm:structural} applies to every graph $G$ of Section~\ref{subsec:search}, and $\nu_b$ can be $0$ when short cycles share vertices (Example~\ref{ex:chain}). At a vertex $x$ shared by cycles of the support of $H$, replacing $H$ by its routing graph can have two effects, depending on the order of the ports of $x$: two simple cycles with fewer than $P$ vertices in the support of the routing graph that are mapped to cycles through $x$ (Lemma~\ref{lem:routing}) can share edges of the path of ports of $x$, and a further simple cycle with fewer than $P$ vertices can pass through two disjoint subpaths of this path, so that its closed walk in the support of $H$ passes twice through $x$. In the first, an edge of the support of the routing graph lies on two simple cycles with fewer than $P$ vertices, so the sufficient condition for $\nu_b=0$ stated before Theorem~\ref{thm:structural} fails for $b=1$; in the second, the closed walk of such a cycle is not a simple cycle. The hypothesis of Theorem~\ref{thm:short-cycles} excludes both effects. From this hypothesis, the last paragraph of the proof above shows that the closed walk of every simple cycle with fewer than $P$ vertices in the support of the routing graph is a simple cycle and that two different such cycles share no edge, which gives $\nu_b=0$. The condition that every edge of the support of $H$ lies on at most one simple cycle with fewer than $P$ vertices does not exclude these effects: it holds for the input of Example~\ref{ex:chain}, and, for $P>16$, both effects occur in the routing graph of this input for some orders of the ports (paragraph after the example).

Example~\ref{ex:chain} avoids both effects by taking $G=H$, which is possible for inputs of bounded degree: for a simple input $H$ whose incoming and outgoing degrees are at most a constant $D$, a variant of Algorithm~\ref{alg:driver} can use $G=H$ and $s'=s$ in line~\ref{line:routing} and test the guards for this $D$ in line~\ref{line:guardtest}. The guards then hold for all sufficiently large $N$, and Corollary~\ref{cor:exact} and Theorem~\ref{thm:structural} apply to $H$ itself.

\begin{example}\label{ex:chain}
Let $g\ge2$. Take vertices $u_1,\dots,u_{g+1}$ and $v_1,\dots,v_g$ and, for every $i\le g$, both arcs between each two of the vertices $u_i$, $v_i$, $u_{i+1}$, with arbitrary nonnegative weights. The support is a chain of $g$ triangles in which consecutive triangles share a vertex, so the hypothesis of Theorem~\ref{thm:short-cycles} fails. The vertices $u_2,\dots,u_g$ are cut vertices, so every simple cycle of the support lies in one triangle, and every edge lies on exactly one simple cycle. Hence no arc is $1$-heavy and $\nu_1=0$. The degrees are at most $4$. With $G=H$ and $s'=s$ as in the preceding paragraph, Theorem~\ref{thm:structural} with $D=4$ and $b=1$ bounds the search on this graph by $O(N\sqrt L)$ for every source and all weights, once $N$ is large enough for the guards to hold.
\end{example}

When $G$ is the routing graph, both effects described before the example can occur on its input. Suppose that $P>16$, which holds for all sufficiently large $g$, and that the ports of some $u_i$ with $2\le i\le g$ lie along their path in the order $u_{i-1},v_i,v_{i-1},u_{i+1}$ of the neighbours. Then the cycles that are mapped to the two triangles at $u_i$ share an edge of this path, and the closed walk $u_iu_{i-1}v_{i-1}u_iu_{i+1}v_iu_i$ is the image of a simple cycle with at most $16$ vertices. The arc from the port of $v_{i-1}$ assigned to $u_i$ to the port of $u_i$ assigned to $v_{i-1}$ is then $1$-heavy for all weights: in this cycle and in the cycle mapped to the triangle $u_{i-1}v_{i-1}u_i$, the zero-weight arc of the path of ports into the head of this arc and the rest of the cycle form a diamond whose losing arc is this arc, since the arm of one zero-weight arc has the smaller triple $\tau$. If every $u_i$ with $2\le i\le g$ has this order of ports, then the cycles mapped to the triangles $u_{i-1}v_{i-1}u_i$ with even $i\le g$ are pairwise vertex-disjoint and each competes at such a $1$-heavy arc, so $\nu_1\ge\lfloor g/2\rfloor$. Since $N=6g$, the bound $O(N\sqrt L+\nu_1L)$ of Theorem~\ref{thm:structural} is then of order $NL$.

%% file: intermediate.tex
\section{Few DFS exceptions}\label{sec:intermediate}

This section proves Theorem~\ref{thm:dfs-exceptions}. The algorithm is Algorithm~\ref{alg:dfs-driver}. It selects an integer budget $q$ and a set $X$ of exceptional vertices such that the rest of the graph has a tree decomposition of width at most $k+1$ along the depth-first search tree. It then runs Dijkstra's algorithm on $X$ and on the vertices of a few bags of this decomposition only, and handles the other vertices by elimination along the decomposition. The section is independent of Section~\ref{sec:sparse}.

\subsection{The algorithm}\label{subsec:dfs-driver}

The input is a directed graph $H$ with $n$ vertices, $m$ arcs, nonnegative real arc weights and a source $s$, together with an integer $k\ge1$. In the budget calculation, $R=n+m+2$ and $\J(n)=\lceil\log_2(n+2)\rceil$ use the original input counts. The vertices outside the weak component of $s$ are unreachable. The algorithm sets their distances to $\infty$, discards them and makes $H$ simple (Section~\ref{sec:model}), in $O(R)$ time. From now on $H$ is simple and its support is connected.

The algorithm computes the depth-first search tree of the support of $H$ that starts at $s$ and visits the neighbours of each vertex in increasing order of their identifiers. Every edge of the support joins a vertex to one of its ancestors in this tree. An arc of $H$ whose endpoints are not joined by a tree edge is a \emph{chord}. Its endpoints are a proper ancestor, the \emph{upper endpoint}, and a descendant, the \emph{lower endpoint}. A chord \emph{crosses the cut above} a vertex $v\ne s$ if its lower endpoint is $v$ or a descendant of $v$ and its upper endpoint is a proper ancestor of $v$. A set $Y$ of vertices is \emph{$k$-feasible} if, for every vertex $v\ne s$, the chords with no endpoint in $Y$ that cross the cut above $v$ have at most $k$ distinct upper endpoints. For a set $Y$ of vertices, $d^+(Y)$ is the number of arcs whose tail is in $Y$, and $I(Y)$ is the number of arcs from $Y$ to its complement. The preprocessing changes neither the tree nor the pairs of endpoints of the chords, so it preserves the $k$-feasible sets of Theorem~\ref{thm:dfs-exceptions}, and it does not increase $d^+(Y)$.

Both routines use an integer \emph{budget} $q\ge1$. The first, $\Select(q)$, returns a $k$-feasible set. Given such a set $X$, $\Solve(X,q)$ returns the distances from $s$. Sections~\ref{subsec:select} and~\ref{subsec:solve} give the two subroutines and prove the following guarantees.

\begin{lemma}\label{lem:select}
For every integer $q\ge1$, $\Select(q)$ returns a $k$-feasible set $X_q$ in $O_k(R)$ time and space, and every $k$-feasible set $Y$ satisfies
\begin{equation}\label{eq:credit}
 q\size{X_q}+d^+(X_q)\le2(k+1)\bigl(q\size Y+d^+(Y)\bigr).
\end{equation}
\end{lemma}

\begin{lemma}\label{lem:solve}
For every $k$-feasible set $X$ and every integer $q\ge1$, $\Solve(X,q)$ returns the distances from $s$ in $O_k(R)$ space and in time
\[
 O_k\bigl(R(1+q)+(2+\size X+I(X)/q)\,\J(n)\bigr).
\]
\end{lemma}

The budget balances the costs in Lemma~\ref{lem:solve}. The routine $\Solve$ keeps at most $\size X+O_k(1+I(X)/q)$ vertices in a priority queue, and it processes each remaining part of the graph at most $q+O(k)$ times. A larger budget lowers the bound on the queue and raises the bound on the number of times a part is processed. The algorithm knows neither the cheapest feasible set nor the right budget. Lemma~\ref{lem:select} makes the selected set competitive with every feasible set at the same budget. Algorithm~\ref{alg:dfs-driver} therefore tries the budgets $1,2,4,\dots$, computes for each the time bound $\Psi$ of Lemma~\ref{lem:solve} for the selected set, and keeps the best pair. It stops when $Rq$ reaches the best bound found, because every later bound exceeds $Rq$.

\begin{algorithm}[htbp]
\caption{Shortest paths with DFS exceptions}\label{alg:dfs-driver}
\KwIn{directed graph $H$ with nonnegative real arc weights; source $s$; an integer $k\ge1$}
\KwOut{the distances from $s$ and a shortest-path tree}
$\dist(v)\gets\infty$ for every vertex $v$ outside the weak component of $s$; discard these vertices\;
remove the loops; for each ordered pair of vertices keep one arc of least weight\;
compute the depth-first search tree from $s$, the chords and their endpoints\;
$q\gets1$;\quad $\Psi^*\gets\infty$\;\label{line:budget-init}
\While{$Rq<\Psi^*$}{\label{line:budget-loop}
  $X\gets\Select(q)$\Rem*{Algorithm~\ref{alg:select}}
  $\Psi\gets R(1+q)+\bigl(2+\size X+\lceil I(X)/q\rceil\bigr)\J(n)$\Rem*{the time bound of Lemma~\ref{lem:solve}}\label{line:score}
  \lIf{$\Psi<\Psi^*$}{$(\Psi^*,X^*,q^*)\gets(\Psi,X,q)$}
  $q\gets2q$\;
}
$\dist\gets\Solve(X^*,q^*)$ on the remaining vertices\Rem*{Algorithm~\ref{alg:solve}}\label{line:final-solve}
\Return $\dist$ and the tree of first discoveries along the tight arcs\Rem*{Section~\ref{sec:model}}
\end{algorithm}

\begin{proof}[Proof of Theorem~\ref{thm:dfs-exceptions}]
The output is correct by Lemma~\ref{lem:solve} and Section~\ref{sec:model}. Write $\Psi_q$ for the value computed in line~\ref{line:score} at budget $q$, and $\Psi^*$, $q^*$ for the final values. Line~\ref{line:final-solve} costs $O_k(\Psi^*)$ by Lemma~\ref{lem:solve}.

\emph{The loop.} Since $2+\size X+I(X)\le R$, we have $\Psi_1\le R(2+\J(n))$, so the loop ends with $q\le2(2+\J(n))$. Every budget $q$ for which the body of the loop runs satisfies $Rq<\Psi^*$: for $q\le q^*$ because $Rq\le Rq^*<\Psi_{q^*}=\Psi^*$, and for $q>q^*$ by the test in line~\ref{line:budget-loop}. So the body runs at most $1+\log_2(\Psi^*/R)$ times. One run costs $O_k(R)$ by Lemma~\ref{lem:select}, and $\Psi^*\ge2R$, so the loop costs $O_k(\Psi^*)$. All integers involved are bounded by a polynomial in $R$. Since $q$ is a power of two, line~\ref{line:score} needs only shifts, additions and one multiplication by $\J(n)$, which takes $O(\log R)$ additions.

\emph{Comparison with a feasible set.} Let $Y$ be $k$-feasible, and let $q_Y$ be the power of two with $\max\{1,\sqrt{d^+(Y)\J(n)/R}\}\le q_Y<2\max\{1,\sqrt{d^+(Y)\J(n)/R}\}$. Then
\begin{equation}\label{eq:balance}
 Rq_Y\le2R+2\sqrt{R\,d^+(Y)\J(n)}\qquad\text{and}\qquad d^+(Y)\J(n)/q_Y\le R+\sqrt{R\,d^+(Y)\J(n)}.
\end{equation}
If the body of the loop ran with $q=q_Y$, then $\Psi^*\le\Psi_{q_Y}$. Since $I(X)\le d^+(X)$, inequality \eqref{eq:credit} gives $\size{X_q}+I(X_q)/q\le2(k+1)(\size Y+d^+(Y)/q)$, hence
\[
 \Psi_{q_Y}\le R(1+q_Y)+\bigl(3+2(k+1)(\size Y+d^+(Y)/q_Y)\bigr)\J(n).
\]
If it did not, the loop ended at some budget $q\le q_Y$ with $\Psi^*\le Rq\le Rq_Y$. In both cases \eqref{eq:balance} gives
\[
 \Psi^*=O_k\bigl(R+(1+\size Y)\J(n)+\sqrt{R\,d^+(Y)\J(n)}\bigr).
\]
The total running time is $O_k(R+\Psi^*)$, and $\J(n)\le3\log_2n$ for $n\ge2$. Taking the minimum over $Y$ proves the time bound. The space bound follows from the two lemmas, since the algorithm stores only the best set found so far.
\end{proof}

\subsection{Selecting the exceptional set}\label{subsec:select}

$\Select(q)$ gives every vertex $v$ an initial \emph{credit} $q+d^+(v)$, where $d^+(v)=d^+(\{v\})$. This is the price of making $v$ exceptional at budget $q$. It processes the cuts from the leaves upward. A chord is \emph{live} while it has no selected endpoint. Whenever the live chords collected so far at a cut have $k+1$ distinct upper endpoints, which we call a \emph{conflict}, it takes one live chord for each of these endpoints, lowers the credits of all endpoints of these chords by the same amount, and selects the vertices whose credit reaches zero. The selected vertices form the output. This rule has the form of the one by which Bar-Yehuda and Even~\cite{BarYehudaEven1981} approximate a vertex cover of minimum weight, where the two endpoints of an uncovered edge lose the smaller of their weights.

For every vertex $v\ne s$ the procedure maintains a collection $\calL(v)$ of lists of chords. Each list has a key, the common upper endpoint of its chords, and different lists of $\calL(v)$ have different keys. The lists passed to $\Add$ hold chords that cross the cut above $v$.

\begin{algorithm}[htbp]
\caption{$\Select(q)$}\label{alg:select}
\KwIn{budget $q\ge1$; the tree, the chords and $k$ are those of Algorithm~\ref{alg:dfs-driver}}
\KwOut{a $k$-feasible set of vertices}
\lForEach{vertex $v$}{$\kappa(v)\gets q+d^+(v)$;\quad $v$ is not selected}
\ForEach{vertex $v\ne s$ in postorder}{\label{line:postorder}
  $\calL(v)\gets\varnothing$\;
  \ForEach{child $c$ of $v$ and each list $\ell\in\calL(c)$ whose key is not $v$}{
    $\Add(v,\ell)$\;\label{line:add-child}
  }
  \ForEach{chord $f$ with lower endpoint $v$}{
    $\Add(v,\langle f\rangle)$\Rem*{a list with the single chord $f$}\label{line:add-own}
  }
}
\Return the set of selected vertices\;
\BlankLine
\Proc{$\Add(v,\ell)$}{
  \leIf{$\calL(v)$ has a list $\ell'$ with the key of $\ell$}{append $\ell$ to $\ell'$}{insert $\ell$ into $\calL(v)$}
  \Repeat{$\size{\calL(v)}\le k$}{\label{line:repeat}
    \ForEach{list $\ell'\in\calL(v)$}{
      remove chords from the front of $\ell'$ while its first chord has a selected endpoint\;\label{line:prune}
      \lIf{$\ell'$ is empty}{delete $\ell'$ from $\calL(v)$}
    }
    \If(\Rem*[f]{a conflict at the cut above $v$}){$\size{\calL(v)}=k+1$}{\label{line:conflict}
      $E\gets$ the set of endpoints of the first chords of the $k+1$ lists\;
      $\delta\gets\min_{x\in E}\kappa(x)$\;
      \ForEach{$x\in E$}{
        $\kappa(x)\gets\kappa(x)-\delta$\;
        \lIf{$\kappa(x)=0$}{select $x$}\label{line:selectx}
      }
    }
  }
}
\end{algorithm}

For example, let $k=1$, and let two live chords with upper endpoints $a_1\ne a_2$ and lower endpoints $x_1,x_2$ be the first chords of the two lists at a cut. This is a conflict with $E=\{a_1,x_1,a_2,x_2\}$. If the credits of these vertices are $5,2,3,4$, then $\delta=2$, the credits become $3,0,1,2$, and $x_1$ is selected. The next pass of the loop of line~\ref{line:repeat} removes the first chord from the list with key $a_1$, and no conflict remains unless this list has another live chord.

\begin{proof}[Proof of Lemma~\ref{lem:select}]
\emph{Invariant.} A chord with lower endpoint $x$ and upper endpoint $a$ is inserted at $x$ in line~\ref{line:add-own} and moves upward in line~\ref{line:add-child} until its list reaches the child of $a$ on the path to $x$; the parent of that child is $a$, the key of the list, so the list is not passed on. Hence the chords in the lists of $\calL(v)$ are chords that cross the cut above $v$, and every chord that crosses this cut and is not in these lists has been removed in line~\ref{line:prune}, with a selected endpoint. Each call of $\Add$ adds at most one list to at most $k$ lists and ends with at most $k$ lists, each of which begins with a live chord.

\emph{Feasibility.} When the loop of line~\ref{line:postorder} has finished with $v$, the live chords that cross the cut above $v$ lie in the at most $k$ lists of $\calL(v)$, so they have at most $k$ distinct upper endpoints. Later selections only remove live chords. Hence the final set of selected vertices is $k$-feasible.

\emph{The inequality.} Number the conflicts of line~\ref{line:conflict} and let $E_i$ and $\delta_i$ be the set $E$ and the amount $\delta$ of the $i$th conflict. The vertices of $E_i$ are not selected, so $\delta_i>0$, a credit never becomes negative, and the conflict selects at least one vertex. Let $Y$ be $k$-feasible. The $k+1$ first chords of the $i$th conflict cross one cut and have distinct upper endpoints, so one of them has an endpoint in $Y$; that is, $E_i$ meets $Y$. The total amount subtracted from the credit of a vertex is at most its initial credit, so
\[
 \sum_i\delta_i\le\sum_{y\in Y}\ \sum_{i:\,y\in E_i}\delta_i\le\sum_{y\in Y}\bigl(q+d^+(y)\bigr)=q\size Y+d^+(Y).
\]
The credit of every vertex of $X_q$ has been used up, and $\size{E_i}\le2(k+1)$, so
\[
 q\size{X_q}+d^+(X_q)\le\sum_i\size{E_i}\,\delta_i\le2(k+1)\sum_i\delta_i ,
\]
which proves \eqref{eq:credit}.

\emph{Running time.} There are at most $k(n-1)+m$ calls of $\Add$, since a child passes at most $k$ lists. Finding the list with a given key among at most $k+1$ lists and appending take $O(k)$ time, with lists that store pointers to both ends. One iteration of the loop of line~\ref{line:repeat} takes $O(k)$ time apart from the removals in line~\ref{line:prune}. Each chord is removed at most once, and every iteration after the first in a call follows a conflict, of which there are at most $n$. The total time is $O_k(R)$, and so is the space.
\end{proof}

\subsection{The distance routine}\label{subsec:solve}

For a $k$-feasible set $X$, the routine $\Solve(X,q)$ runs Dijkstra's algorithm on $X$ and on the vertices of a few bags of a tree decomposition of $H-X$ only. We first construct the decomposition.

Let $X$ be $k$-feasible. For a vertex $v\ne s$ let $\up(v)$ be the set of upper endpoints of the chords with no endpoint in $X$ that cross the cut above $v$; thus $\size{\up(v)}\le k$. Define the \emph{bags}
\[
 \bag(v)\coloneqq\bigl(\{v,\text{the parent of }v\}\cup\up(v)\bigr)\setminus X\quad(v\ne s),\qquad\bag(s)\coloneqq\{s\}\setminus X .
\]
A bag has at most $k+2$ vertices. The bags are indexed by the nodes of the depth-first search tree, and two bags are \emph{adjacent} when their nodes are.

\begin{lemma}\label{lem:bags}
The bags form a tree decomposition of $H-X$ along the depth-first search tree: every vertex $v\notin X$ lies in $\bag(v)$, the two endpoints of every arc of $H-X$ lie in a common bag, and for every vertex the nodes whose bags contain it induce a connected subtree. Each of these nodes is the vertex itself or one of its descendants. All bags can be computed in $O_k(R)$ time.
\end{lemma}
\begin{proof}
An arc of $H-X$ on a tree edge joins a vertex $v$ to its parent, and both lie in $\bag(v)$. Every other arc of $H-X$ is a chord with no endpoint in $X$. It crosses the cut above its lower endpoint $x$, so its upper endpoint lies in $\up(x)$ and both endpoints lie in $\bag(x)$.

Let $a\notin X$. The bags that contain $a$ are $\bag(a)$, the bags of the children of $a$, and the bags $\bag(v)$ with $a\in\up(v)$. If $a\in\up(v)$ because of a chord with lower endpoint $x$, then this chord crosses the cut above every vertex on the tree path from $x$ up to the child of $a$ on this path, so $a$ lies in the bags of all these vertices. The nodes whose bags contain $a$ are therefore connected, and each of them is $a$ or a descendant of $a$.

A postorder traversal computes $\up(v)$ from the sets $\up(c)$ of the children $c$ of $v$, without the element $v$, and from the chords with lower endpoint $v$ and no endpoint in $X$. Each set has at most $k$ elements, so this takes $O(k)$ time per child and per chord.
\end{proof}

$\Solve$ marks some bags. The \emph{terminals} are the vertices of $X$ and the vertices of the marked bags; let $Z$ be the set of terminals. A \emph{piece} is a connected component of the unmarked nodes of the tree. A vertex outside $Z$ lies only in unmarked bags, and their nodes are connected by Lemma~\ref{lem:bags}, so they belong to one piece $\Pi$; the vertex is an \emph{interior vertex} of $\Pi$, and $\Pi^\circ$ denotes the set of interior vertices of $\Pi$.

Dijkstra's algorithm runs on the terminals, and a piece is processed once for every terminal with an arc into its interior and once more at the end. Give every vertex $v\notin X$ a \emph{mass}, the number of arcs from $X$ to $v$. Line~\ref{line:mass-mark} of Algorithm~\ref{alg:solve} marks the bag of a node as soon as the component of this node among the unmarked nodes of its subtree has mass more than $q$, and line~\ref{line:branching} marks the bags of the branching nodes of the smallest subtree that contains all marked nodes. This marking bounds both numbers: at most $2(1+I(X)/q)$ bags are marked, and at most $q+2(k+2)$ terminals have an arc into the interior of a piece (proof of Lemma~\ref{lem:solve}).

For a piece $\Pi$ and a \emph{seed} function $\varsigma$ on $\Pi^\circ$ with values in $[0,\infty]$, the subroutine $\Interior(\Pi,\varsigma)$ returns the function
\begin{equation}\label{eq:interior}
 \eta(y)=\min_{x\in\Pi^\circ}\bigl(\varsigma(x)+\dist_{H[\Pi^\circ]}(x,y)\bigr)\qquad(y\in\Pi^\circ),
\end{equation}
where $\dist_{H[\Pi^\circ]}$ is the distance in the subgraph induced on $\Pi^\circ$. A seed that is given only on some vertices has the value $\infty$ on the others. The subroutine eliminates the vertices of $\Pi^\circ$ in postorder, so that every vertex precedes its proper ancestors; this is the \emph{elimination order}. For an interior vertex $y$ put $N_y\coloneqq(\bag(y)\setminus\{y\})\setminus Z$. The vertices of $N_y$ are proper ancestors of $y$ and interior vertices of the piece of $y$.

\begin{algorithm}[htbp]
\caption{$\Solve(X,q)$}\label{alg:solve}
\KwIn{a $k$-feasible set $X$; a budget $q\ge1$; the graph $H$, the tree and the chords are those of Algorithm~\ref{alg:dfs-driver}}
\KwOut{the distances from $s$}
compute $\bag(v)$ for every vertex $v$\Rem*{Lemma~\ref{lem:bags}}
\ForEach(\Rem*[f]{mark bags by mass}){vertex $v$ in postorder}{\label{line:mass}
  $\mu(v)\gets\sum\{\mu(c):\ c\text{ a child of }v\}$\;
  \lIf{$v\notin X$}{$\mu(v)\gets\mu(v)+{}$the number of arcs from $X$ to $v$}
  \lIf{$\mu(v)>q$}{mark $\bag(v)$;\quad $\mu(v)\gets0$}\label{line:mass-mark}
}
mark $\bag(s)$\;
mark every bag whose node has at least three neighbours in the smallest subtree that contains all marked nodes\;\label{line:branching}
$Z\gets X\cup\bigcup\{\bag(v):\ \bag(v)\text{ is marked}\}$\;
compute the pieces, their interior vertices in postorder and the sets $N_y$; group the arcs from $Z$ into interiors by tail and piece\;\label{line:prep}
$\dist(s)\gets0$;\quad $\dist(v)\gets\infty$ for every $v\ne s$;\quad $F\gets$ a Fibonacci heap that holds $Z$ with keys $\dist$\;
\While(\Rem*[f]{Dijkstra's algorithm on $Z$}){$F$ is nonempty}{\label{line:dijkstra}
  $a\gets$ a vertex of least key in $F$; remove $a$ from $F$\;
  \lForEach{arc $(a,z)$ with $z\in Z$}{$\dist(z)\gets\min\{\dist(z),\dist(a)+w(a,z)\}$}
  \ForEach{piece $\Pi$ with an arc from $a$ into $\Pi^\circ$}{\label{line:enter}
    $\eta\gets\Interior\bigl(\Pi,\ x\mapsto\dist(a)+w(a,x)\text{ for the arcs }(a,x)\text{ into }\Pi^\circ\bigr)$\Rem*{Algorithm~\ref{alg:interior}}\label{line:interior-a}
    \ForEach{arc $(y,z)$ with $y\in\Pi^\circ$ and $z\in Z$}{
      $\dist(z)\gets\min\{\dist(z),\eta(y)+w(y,z)\}$\;
    }
  }
}
\ForEach(\Rem*[f]{interior vertices}){piece $\Pi$}{\label{line:final-interior}
  $\eta\gets\Interior\bigl(\Pi,\ x\mapsto\min\{\dist(a)+w(a,x):\ a\in Z,\ (a,x)\text{ an arc}\}\bigr)$\;
  \lForEach{$y\in\Pi^\circ$}{$\dist(y)\gets\eta(y)$}
}
\Return $\dist$\;
\end{algorithm}

A shortest-path tree of a directed graph of constant treewidth can be computed in linear time~\cite{ChaudhuriZaroliagis2000}. Algorithm~\ref{alg:interior} computes the seeded function $\eta$ of \eqref{eq:interior} directly, by eliminating the vertices of $\Pi^\circ$ along the decomposition; Lemma~\ref{lem:interior} proves this. For every ordered pair $x,x'$ with $x'\in N_x$ or $x\in N_{x'}$, it stores a shortcut length $\omega(x,x')$, initially the arc weight or $\infty$ if there is no such arc.

\begin{algorithm}[htbp]
\caption{$\Interior(\Pi,\varsigma)$}\label{alg:interior}
\KwIn{a piece $\Pi$; a seed $\varsigma(y)\in[0,\infty]$ for every $y\in\Pi^\circ$; the sets $N_y$ and the order of $\Pi^\circ$ are those of Algorithm~\ref{alg:solve}}
\KwOut{the function $\eta$ of \eqref{eq:interior}}
\ForEach{$y\in\Pi^\circ$ and $x\in N_y$}{
  $\omega(y,x)\gets w(y,x)$;\quad $\omega(x,y)\gets w(x,y)$\Rem*{$\infty$ if the arc is absent}
}
\ForEach{$y\in\Pi^\circ$ in elimination order}{\label{line:eliminate}
  \lForEach{$x\in N_y$}{$\varsigma(x)\gets\min\{\varsigma(x),\varsigma(y)+\omega(y,x)\}$}
  \lForEach{ordered pair $x\ne x'$ in $N_y$}{$\omega(x,x')\gets\min\{\omega(x,x'),\omega(x,y)+\omega(y,x')\}$}\label{line:omega}
}
\ForEach{$y\in\Pi^\circ$ in reverse elimination order}{\label{line:back}
  $\eta(y)\gets\min\bigl(\{\varsigma(y)\}\cup\{\eta(x)+\omega(x,y):\ x\in N_y\}\bigr)$\;\label{line:eta}
}
\Return $\eta$\;
\end{algorithm}

\begin{lemma}\label{lem:interior}
$\Interior(\Pi,\varsigma)$ returns the function $\eta$ of \eqref{eq:interior}. One call takes $O_k(1)$ time per node of $\Pi$ and per arc with an endpoint in $\Pi^\circ$.
\end{lemma}
\begin{proof}
\emph{Stored pairs.} The bags of $\Pi$, restricted to $\Pi^\circ$, form a tree decomposition of $H[\Pi^\circ]$ by Lemma~\ref{lem:bags}. Let $x\ne y$ be vertices of $\Pi^\circ$ that share a bag, and let $y$ precede $x$ in the elimination order. By Lemma~\ref{lem:bags} the common bag belongs to a common descendant of $x$ and $y$, so one of the two is an ancestor of the other, and $x$ is a proper ancestor of $y$. The nodes whose bags contain $x$ are connected and include $x$ and this descendant, hence also $y$. So $x\in N_y$. In particular, two vertices of $N_y$ share $\bag(y)$, so the value $\omega(x,x')$ changed in line~\ref{line:omega} is one of the stored values.

\emph{Forward pass (lines~\ref{line:eliminate}--\ref{line:omega}).} When the loop of line~\ref{line:eliminate} reaches $y$, the following holds for all $x,x'$ that do not precede $y$. Every stored value $\omega(x,x')$ is $\infty$ or the length of a walk of $H[\Pi^\circ]$ from $x$ to $x'$ whose inner vertices precede $y$, and every value $\varsigma(x)$ is $\infty$ or a seed value plus the length of a walk of $H[\Pi^\circ]$ to $x$ whose other vertices precede $y$. If distinct vertices $x,x'$ are joined by a path of $H[\Pi^\circ]$ from $x$ to $x'$ whose inner vertices precede $y$, then $x$ and $x'$ share a bag and $\omega(x,x')$ is the least length of such a path. Moreover $\varsigma(x)$ is the least value of a seed plus the length of a path to $x$ whose other vertices precede $y$. All of this holds for the first vertex: no vertex precedes it, the stored values are then the arc weights, $\varsigma$ is the seed, and the endpoints of an arc share a bag by Lemma~\ref{lem:bags}. The two inner loops preserve this: a shortest such path through $y$ splits at $y$ into two paths of the previous kind, whose other endpoints share a bag with $y$ and therefore lie in $N_y$, and a path that avoids $y$ is itself of the previous kind. The values $\varsigma(y)$ and $\omega(x,y)$ for $x\in N_y$ do not change after $y$ is processed, since only proper descendants of $y$, which precede $y$, change them.

\emph{Backward pass (lines~\ref{line:back}--\ref{line:eta}).} Consider a seeded path that attains $\eta(y)$ in \eqref{eq:interior}. If all its vertices other than $y$ precede $y$, its value is $\varsigma(y)$. Otherwise let $x$ be its last vertex that follows $y$; the part from $x$ to $y$ has inner vertices that precede $y$, so $x\in N_y$ and the value is $\eta(x)+\omega(x,y)$. Line~\ref{line:eta} takes the minimum over these cases, and $\eta(x)$ is known because $x$ follows $y$.

\emph{Time.} A call initializes and processes $O(k^2)$ values per vertex of $\Pi^\circ$ and reads the arcs of $H[\Pi^\circ]$.
\end{proof}

\begin{proof}[Proof of Lemma~\ref{lem:solve}]
\emph{Correctness.} Consider the directed graph on $Z$ that has an arc $(a,z)$ of weight $w(a,z)$ for every arc of $H$ between terminals, and, for every piece $\Pi$, an arc $(a,z)$ of weight $\min\{w(a,x)+\dist_{H[\Pi^\circ]}(x,y)+w(y,z)\}$ over the arcs $(a,x)$ and $(y,z)$ of $H$ with $x,y\in\Pi^\circ$. The weight of each of its arcs is the length of a walk of $H$. Conversely, split a shortest path of $H$ from $s$ to a terminal at its terminals. Two consecutive terminals are joined by an arc, or by a subpath whose inner vertices are outside $Z$. Consecutive inner vertices are joined by an arc of $H-X$, so they share a bag, and they are interior to the same piece. Hence the distances from $s$ in this graph are the distances of $H$ on $Z$. The loop of line~\ref{line:dijkstra} is Dijkstra's algorithm on this graph: when $a$ is removed from $F$, the loop relaxes the arcs out of $a$, and \eqref{eq:interior} gives the arcs through each piece. Finally, a shortest path to an interior vertex $y$ of a piece leaves its last terminal $a$ by an arc $(a,x)$ and continues inside the interior, so line~\ref{line:final-interior} computes $\dist(y)$; the source is a terminal, so such a terminal exists.

\emph{Marked bags.} The total mass is $I(X)$. After the loop of line~\ref{line:mass}, for every unmarked node $v$, $\mu(v)$ is the mass of the component of $v$ among the unmarked nodes of the subtree of $v$. Each mark of line~\ref{line:mass-mark} accounts for more than $q$ units of mass that no other mark accounts for, so there are at most $I(X)/q$ such marks, and every component of the unmarked nodes has mass at most $q$, the value $\mu$ of its top node. The smallest subtree containing $i$ marked nodes has fewer than $i$ nodes with at least three neighbours. Hence at most $2(1+I(X)/q)$ bags are marked, and
\[
 \size Z\le\size X+2(k+2)\bigl(1+I(X)/q\bigr).
\]
Line~\ref{line:branching} only splits components, so every piece has mass at most $q$.

\emph{Marked bags next to a piece.} A piece is adjacent to at most two marked nodes. Suppose that three marked nodes are adjacent to nodes $c_1,c_2,c_3$ of a piece $\Pi$. The paths between these marked nodes pass through $\Pi$, and the node of $\Pi$ at which the tree paths between $c_1,c_2,c_3$ meet has three neighbours in the subtree of line~\ref{line:branching}. Its bag is marked, a contradiction.

\emph{Terminals entering a piece.} Let $(a,x)$ be an arc from a terminal $a$ to an interior vertex $x$ of $\Pi$. If $a\in X$, the arc contributes one unit to the mass of $x$, and the node of $x$ belongs to $\Pi$ because $\bag(x)$ contains $x$. So at most $q$ arcs from $X$ enter $\Pi^\circ$. If $a\notin X$, the arc belongs to $H-X$, so some bag contains $a$ and $x$ by Lemma~\ref{lem:bags}, and its node is in $\Pi$. The vertex $a$ also lies in a marked bag. The nodes whose bags contain $a$ are connected, so the first node outside $\Pi$ on a path among them to a marked node is a marked node adjacent to $\Pi$. So $a$ lies in one of at most two bags. Altogether at most $q+2(k+2)$ terminals have an arc into $\Pi^\circ$.

\emph{Running time.} Since $\bag(s)$ is marked, the smallest subtree that contains all marked nodes consists of the nodes with a marked node in their subtree, and a node other than $s$ has at least three neighbours in it exactly when at least two of its children belong to it. One postorder traversal therefore executes line~\ref{line:branching}. Line~\ref{line:prep} takes $O_k(R)$ time: one more traversal yields the pieces and lists their interior vertices in postorder, the sets $N_y$ are read off the bags, and sorting the arcs from terminals into interiors by tail and piece gives the lists used in line~\ref{line:enter}. Let the volume of a piece be the number of its nodes plus the number of arcs with an endpoint in its interior. The volumes sum to at most $n+2m$. A piece is passed to $\Interior$ once for each terminal with an arc into its interior and once in line~\ref{line:final-interior}, so at most $q+2k+5$ times, and each time its arcs to terminals are read once. By Lemma~\ref{lem:interior} this costs $O_k(R(1+q))$ in total. The heap performs $\size Z$ removals at cost $O(\log\size Z)=O(\J(n))$ each, and its other operations take constant amortized time~\cite{FredmanTarjan1987}. The space is $O_k(R)$, since the working storage of $\Interior$ is reused.
\end{proof}

\subsection{The density \texorpdfstring{$n\sqrt{\log\log n}$}{n sqrt(log log n)}}\label{subsec:focal}

The following corollary concerns the density $m=\Theta(n\sqrt{\log\log n})$, the case $\alpha=1/2$ of~\cite[Corollary~1.2]{GZ}, whose bound holds for all inputs of this density and carries the factor $(\log\log n)^{5/8}$.

\begin{corollary}\label{cor:focal}
Let $k$ be fixed and $m=\Theta(n\sqrt{\log\log n})$. If some $k$-feasible set $Y$ has $\size Y=O(n/\sqrt{\log n})$ and $d^+(Y)=O(n/\sqrt{\log\log n})$, then Algorithm~\ref{alg:dfs-driver} runs in $O_k(n\sqrt{\log n})$ time.
\end{corollary}
\begin{proof}
In the bound of Theorem~\ref{thm:dfs-exceptions}, the terms $n+m$, $(1+\size Y)\log n$ and $\sqrt{(n+m)\,d^+(Y)\log n}$ are all $O(n\sqrt{\log n})$.
\end{proof}

\begin{example}\label{ex:path-square}
Let $g\coloneqq\lceil\sqrt{\log\log(n+8)}\,\rceil$ and let $n$ be large. On the vertices $1,\dots,n$ take both arcs between $i$ and $j$ whenever $1\le\size{i-j}\le2$, and an arc from every vertex $i>n/2$ to every vertex $a\le g$, with arbitrary nonnegative weights and source $1$. The graph is strongly connected and has $\Theta(n\sqrt{\log\log n})$ arcs. The depth-first search visits $1,2,\dots,n$ in this order, because the least unvisited neighbour of $i$ is $i+1$. The chords are the arcs between $i$ and $i+2$ and the arcs into $\{1,\dots,g\}$ from the upper half. The chords with no endpoint in $Y\coloneqq\{1,\dots,g\}$ cross the cut above $v$ with upper endpoints in $\{v-2,v-1\}$, so $Y$ is $2$-feasible, with $\size Y=g$ and $d^+(Y)\le4g$. Theorem~\ref{thm:dfs-exceptions} with $k=2$ gives $O(n+m)$ time. Without exceptional vertices, $k$ would have to be at least $g+2$, because at the cut above a vertex $v$ with $n/2<v<n$ the chords have $g+2$ upper endpoints; since $g$ grows with $n$, no fixed $k$ suffices.
\end{example}

\subsection{Exceptional heads}\label{subsec:heads}

No budget is needed for the following one-sided notion of exception, in which a set exempts only the chords whose head it contains. Let $h_k$ be the least size of a set $Y$ such that, for every vertex $v\ne s$, the chords whose head is outside $Y$ and that cross the cut above $v$ have at most $k$ distinct upper endpoints. The following variant of Algorithm~\ref{alg:dfs-driver} replaces its lines~\ref{line:budget-init}--\ref{line:final-solve} and has no loop over budgets.
\begin{enumerate}[label=\arabic*.]
\item Run $\Select$ without credits: a chord is live while its head is not selected, line~\ref{line:prune} removes the chords that are not live, and a conflict selects the heads of its $k+1$ first chords. Let $X'$ be the set of selected vertices, and let $H'$ be $H$ without the chords whose head is in $X'$.
\item Let $\up'(v)$ consist of the upper endpoints of the chords of $H'$ that cross the cut above $v$, and put $\bag'(v)\coloneqq\{v,\text{the parent of }v\}\cup\up'(v)$ for $v\ne s$ and $\bag'(s)\coloneqq\{s\}$.
\item Mark $\bag'(s)$ and $\bag'(v)$ for every $v\in X'$, and run Algorithm~\ref{alg:solve} from line~\ref{line:branching} on, with the bags $\bag'$ and with $X'$ in place of $X$.
\end{enumerate}

\begin{proposition}\label{prop:heads}
For every fixed $k\ge1$, this algorithm computes the distances from $s$ in $O_k\bigl(n+m+(1+h_k)\log(2+h_k)\bigr)$ time and $O_k(n+m)$ space.
\end{proposition}
\begin{proof}
The invariant in the proof of Lemma~\ref{lem:select} is unchanged, so the chords whose head is outside $X'$ cross every cut with at most $k$ distinct upper endpoints. Let $Y$ attain $h_k$. Every conflict has a chord whose head is in $Y$, and the heads of the chords of different conflicts are different, because a conflict selects the heads of its first chords, which were not selected before. So there are at most $h_k$ conflicts and $\size{X'}\le(k+1)h_k$.

The sets $\bag'(v)$ form a tree decomposition of $H'$ with bags of at most $k+2$ vertices; the proof is that of Lemma~\ref{lem:bags}. We have $X'\subseteq Z$ and $\size Z=O_k(1+h_k)$. Every arc of $H$ outside $H'$ has its head in $X'\subseteq Z$. An arc from a terminal into the interior of a piece is therefore an arc of $H'$, and, since every terminal lies in a marked bag (step~3 marks the bags of $X'$), the argument for $a\notin X$ in the proof of Lemma~\ref{lem:solve} shows that at most $2(k+2)$ terminals have such an arc. The arcs between interior vertices of a piece also belong to $H'$, so Lemma~\ref{lem:interior} applies, and the correctness proof is unchanged. Every piece is passed to $\Interior$ at most $2k+5$ times, for a total of $O_k(n+m)$ time, and the heap costs $O(\size Z\log(2+\size Z))$. Steps 1 and 2 and lines~\ref{line:branching}--\ref{line:prep} of Algorithm~\ref{alg:solve} take $O_k(n+m)$ time and space by the proofs of Lemmas~\ref{lem:select}, \ref{lem:bags} and~\ref{lem:solve}, respectively.
\end{proof}

On the graphs of Example~\ref{ex:path-square}, $h_2\le g$, since the chords with heads outside $\{1,\dots,g\}$ are arcs between $i$ and $i+2$.

%% file: conclusion.tex
\section{Conclusion}\label{sec:conclusion}

Our algorithms remove the factors $(\log\log n)^{1/4}$ and $(\log\log n)^{5/8}$ from the bounds of~\cite{GZ} on two main families of inputs. For $m=O(n)$ the family consists of the inputs in which no two simple cycles of the support with fewer than $P$ vertices share a vertex, and for every fixed $c>0$ a uniformly random support with $\lfloor cn\rfloor$ edges has this property with probability at least $1-n^{-1+o(1)}$. For $m=\Theta(n\sqrt{\log\log n})$ and a fixed $k$, the family consists of the inputs in which some $k$-feasible set $Y$ has $\size Y=O(n/\sqrt{\log n})$ and $d^+(Y)=O(n/\sqrt{\log\log n})$; the algorithm is given $k$ but not this set. Whether $O(n\sqrt{\log n})$ time is achievable for all inputs with $m=O(n)$, or for all inputs with $m=\Theta(n\sqrt{\log\log n})$, remains open. The algorithm of Theorem~\ref{thm:typical-sparse}, like that of~\cite{GZ}, runs the recursive search only on very large inputs, here only when $n+2m+2>2^{4095}$, and runs Dijkstra's algorithm otherwise. The algorithm of Theorem~\ref{thm:dfs-exceptions} has no such threshold.

%% file: import.tex
\section{How the new preparation step fits the earlier proof}\label{app:import}

Lemma~\ref{lem:contract} restates~\cite[Theorem~4.1]{GZ} with the hypotheses (F) and (F$'$) of~\cite[Section~4.1]{GZ} on the bounded searches of $\operatorname{Forest}$ replaced by (A1)--(A4). In~\cite{GZ} the scales are written $(t,k)$; here $(t,k)=(T,P)$, and we write $T$ and $P$ also when we quote~\cite{GZ}. The search uses the explicit worklists, the base search tree and the full outgoing scans of~\cite[Section~5.2]{GZ}, whose numerical requirements are the guards \eqref{eq:guards}. The hypotheses of~\cite[Theorem~4.1]{GZ} other than (F) and (F$'$) hold as in~\cite[Section~5.2]{GZ}: the guards give $P\ge2$; with $P\le T/16$ from \eqref{eq:scales} they give the capacity guards $(3P+1)M_j\le A_{j-1}$ and $H_j+2H_{j-1}+PA_j<2H_j$ of~\cite[(12)]{GZ}; the top call has one source, and item~3 bounds the sources of the recursive calls. The search of this paper contains no capped invocation of~\cite[Section~4.3]{GZ}, so no outward stop interrupts a call, and every call is ordinary in the sense of~\cite[Section~4.2]{GZ}, as the conclusion of that theorem requires. The proof of that theorem~\cite[Appendix~D]{GZ} maintains six invariants (J1)--(J6) of the loop of a call. We list the places where this proof, the counting arguments used in Section~\ref{subsec:time} and the proof of~\cite[Lemma~5.1]{GZ} refer to the preparation step.

\begin{enumerate}[label=\arabic*.]
\item \emph{The three forest properties}~\cite[(8)--(10)]{GZ}. The first is (A1). The other two follow from the first and (W3) by the argument given after property (F) in~\cite[Section~4.1]{GZ}. The proof uses the three properties for the invariants (J1)--(J6) at the start of the loop, for the input condition (I2) of a recursive call, for maintaining (J3) during the loop, for the completeness of the output $W_{\mathrm{new}}$, and to show that $W$ contains every target when there is no group. It derives them from (F); here they hold by (A1), and they persist after the step, since $P_S$ and $W$ do not change and complete vertices remain complete (W3).
\item \emph{The set $W$.} The proof uses $W\subseteq\calT(B,S)$ in the argument for $W_{\mathrm{new}}$ and $\size W\le P\size Q\le PA_j$ in the bound $\size U<2H_j$. Both are (A2).
\item \emph{The groups.} The proof uses that the groups are disjoint nonempty sets of sources that cover $P_S$ and have fewer than $3P$ members. It uses this for the bound $\size{S'}\le(3P+1)M_j\le A_{j-1}$ on the sources of a recursive call, where the second inequality is the guard $16(3P+1)\le T$ because $A_{j-1}=TM_j/16$; in the invariants (J2) and (J4), where a pulled member of a group becomes a source of the recursive call and is returned or restored by it; and for the fact that the representatives in the worklist are sources in $P_S$, a fact that the invariants (J3) and (J5) and the record bound use and that gives the bound $a_X<\beta$ of~\cite[Theorem~4.1, item~1]{GZ}, where $a_X$ is the least signature of a source~\cite[Section~4.1]{GZ}. This is (A3). The trees and pieces from which $\operatorname{Forest}$ forms the groups enter the proof of the theorem only where the groups are formed (in the first paragraph of~\cite[Appendix~D]{GZ}) and in its termination argument; here (A3) and the termination of the step assumed in Lemma~\ref{lem:contract} take their place.
\item \emph{Labels and worklists.} The proof uses that an accepted update along an arc from a target with a candidate below $B$ leads to a target. This is the containment argument of~\cite[Lemma~4.2]{GZ}, which needs only (I2) of the input conditions. By (A4) the label changes of the preparation step are updates of this kind, so (W1)--(W3) and the containment remain valid after the step. The proof also uses that the preparation step creates no worklist entry, for the invariants (J3) and (J5) and for the bound $A_j+2DH_j$ on the number of worklist records. This is the second part of (A4).
\item \emph{Property} (F$'$). Apart from the containment in the target, which items 2 and 4 cover, (F$'$) describes the trees of the bounded searches: how they stop, where they meet, and which of their vertices lie outside the target. The proof of the theorem uses it only to construct the groups. The cost analysis of~\cite{GZ} uses it to bound the number of vertices that the searches of one level retain (\cite[Section~4.6]{GZ}, by way of~\cite[Lemma~4.2]{GZ}, and the proof of~\cite[Lemma~5.1]{GZ}); Section~\ref{subsec:time} replaces that bound by $V_X\subseteq U_X$ and (R3) for a fully returning call, and by $\size{V_X}\le P\size{S_X}\le\size{U_X}$ for a partially returning one.
\item \emph{The bound}~\cite[(53)]{GZ}. Section~\ref{subsec:time} uses two counting arguments of~\cite[Appendix~E]{GZ}; this is the first. The proof of the bound on $\sum_X\size{Q_X}$ tracks the occurrences of a vertex in the sets $Q_X$ of nested fully returning calls. It uses that the outputs of the calls of one level are disjoint, that a vertex of $Q_X$ belongs to no group of $X$, that the preparation step creates no worklist entry, that a fully returning call returns its sources and a partially returning one has $T\size S\le\size U$, and that every arc causes at most one insertion from a completed tail; the last fact concerns only the scans of Algorithm~\ref{alg:search} and of the searches of level zero.
\item \emph{The colouring argument of}~\cite[Appendix~E.2]{GZ}. It colours every vertex $v$ of the tree piece of a group by the recursive call whose handled interval~\cite[Section~4.4]{GZ} contains the fixed signature of $v$, or by one auxiliary colour if no such call exists. The fixed signature is $\sig(\mu(v))$ in the notation of~\cite[(13)]{GZ}. It uses that the groups of a call lie in pairwise edge-disjoint connected subgraphs of the support of $G$ and that, in a fully returning call, the vertices of these subgraphs have their fixed signatures in the interval handled by the call. For a target $v$ we have $\mu(v)=\pi(v)$ in the global context of~\cite[Section~4.1]{GZ}, so its fixed signature is $\sig(\pi(v))$, and the targets of a fully returning call are returned by it, so the argument applies to the subgraphs named in Section~\ref{subsec:time}: they are connected, they consist of claimed vertices, which are targets, and the pieces of one spanning tree are edge-disjoint while different components are vertex-disjoint.
\item \emph{The remaining work in the proof of}~\cite[Lemma~5.1]{GZ}. Table~\ref{tab:charges} counts in its own rows the initial representatives, the reselections, the expansions of groups, the scans of the sets $W_{\mathrm{new}}$ and the work that replaces the bounded searches of $\operatorname{Forest}$, and \eqref{eq:inherited} omits the terms of~\cite[(17)]{GZ} that contain $h_{\mathrm{loc}}$ or $T/P$. The rest of that proof bounds the ordinary scans, the merges, conversions and pulled records of the worklists, the restorations of sources, the insertions from completed tails, the base search trees, the sorting of the outgoing lists, the array setup and the per-level arrays of~\cite[Appendix~F.1]{GZ}. It uses that the outputs of the calls of one level are disjoint subsets of the output of the top call, that a partially returning call has $T\size S\le\size U$, that a fully returning call returns its sources and an empty worklist, that a partially returning one returns at most $\size S+D\size U$ records, and that every call returns at least one vertex; these are (R2), (R3) and items 1, 2, 3 and 5 of~\cite[Theorem~4.1]{GZ}, which Lemma~\ref{lem:contract} provides. It also uses that a call at level~$0$ has at most $A_0$ sources (item~3 above), that every arc causes at most one insertion from a completed tail (item~6), and that an insertion costs $O(T)$ (Section~\ref{subsec:search}) because a worklist has at most $A_j+2DH_j$ records. The proofs of items 1 and 3 of~\cite[Theorem~4.1]{GZ}, of this record bound and of the count of pulled records~\cite[Appendix~E.4]{GZ} use items 1 to 4 above. The preparation step enters these parts in no other way.
\end{enumerate}